\documentclass{aa}  

\usepackage{graphicx}
\usepackage{xcolor}
\usepackage{caption}
\usepackage{multirow}
\usepackage{txfonts}
\def\gtsima{$\; \buildrel > \over \sim \;$}
\def\ltsima{$\; \buildrel < \over \sim \;$}
\def\gsim{\lower.5ex\hbox{\gtsima}}
\def\lsim{\lower.5ex\hbox{\ltsima}}

\usepackage{hyperref}
\hypersetup{colorlinks,citecolor=blue}
\usepackage{threeparttable}

\begin{document}

   \title{An extreme Ultraluminous X-ray source X-1 in NGC 5055}


   \author{Samaresh Mondal
		  \thanks{E-mail: smondal@camk.edu.pl (SM)},
          Agata R$\acute{\rm o}\dot{\rm z}$a$\acute{\rm n}$ska,
          Eleonora Veronica Lai,
          \and
          Barbara De Marco
          }

   \institute{Nicolaus Copernicus Astronomical Center, Polish Academy of Sciences,
              ul. Bartycka 18, 00-716 Warsaw, Poland
             }

   \date{Received XXX; accepted YYY}
   
\authorrunning{Mondal et al.}
\titlerunning{ULX NGC 5055 X-1}

 
  \abstract
   {}
   {We analyzed multi-epoch X-ray data of the Ultraluminous X-ray source (ULX) NGC 5055 X-1, with luminosity up to 
   $2.32\times10^{40}\ \rm erg\ s^{-1}$, in order to constrain the physical parameters of the source. }
   {We performed timing and spectral analysis of \emph{Chandra} and XMM-Newton observations. We used spectral models which 
   assume the emission is from an accreting black hole system. We fit the data with a multicolor disk (MCD) combined with a 
   powerlaw (PL) or a thermal Comptonization (NTHCOMP) component, and compared those fits with a slim disk model.}
   {The lightcurves  of the source do not show significant variability. From the hardness ratios (3-10 keV/0.3-3 keV flux) 
   we infer that the source is not spectrally variable. We found that the photon index is 
   tightly, positively correlated with the unabsorbed 0.3-10 keV flux and the hydrogen column density. 
   Furthermore, the temperature emissivity profile indicates a deviation from the standard sub-Eddington thin disk 
   model. The source shows an inverse correlation between luminosity and inner disk temperature in all fitted models.}
  {Our analysis favors the source to be in an ultraluminous soft state. 
  The positive correlations between the photon index and the flux, and between the photon index and the hydrogen column density may suggest the source is 
  accreting at high Eddington ratios 
  and might indicate the presence of a wind.  The inverse luminosity relation with the inner disk temperature for all spectral models may indicate
  that the emission is geometrically beamed by an optically thick outflow.}


   \maketitle
%

\section{Introduction}
Ultraluminous X-ray sources (ULXs) are off-nuclear point sources with isotropic X-ray luminosity in excess of 
$10^{39}\ \rm erg\ s^{-1}$ \citep{fabbiano1989}. 
Due to their high luminosity ULXs were suggested to host an intermediate mass black hole 
\citep[IMBH,][]{colbert1999}; however, the recent discovery of coherent pulsations showed that some 
ULXs contain a neutron star \citep{bachetti14,furst2016,furst2017, 
israel2017a,israel2017b,carpano2018}. This finding was independently confirmed in three sources, through  
fitting with a spectral model which assumes non-isotropic emission from a neutron star - accretion disk system \citep{rozanska2018}.
ULXs are also exciting in the context of gravitational wave studies. Some ULXs have a high mass donor \citep{motch2011, 
motch2014,heida2015,heida2016} that might eventually result in the merging of two compact objects at the end of its 
stellar evolution \citep{mondal2020}.

As most ULXs are extra-galactic sources, the direct measurement of the mass of the compact object is 
extremely difficult. Therefore, we have to rely on indirect methods, like 
BH mass-scaling of the inner disk temperature ($T_{\rm in} \propto M^{-1/4}$) \citep{miller2003}. Most ULX spectra are 
fitted with two components model: a multicolor accretion disk model (hereafter MCD) and a hard powerlaw (hereafter PL) tail. 
Fitting of this model to many ULX spectra results in very cool inner disk temperature, $kT_{\rm in}\sim0.1-0.3$ 
keV and a PL tail $\Gamma \sim 1.5-3$ \citep{kaaret2003,
miller2004a,miller2004b}. If this temperature corresponds to the temperature at the inner disk radius, 
then the inferred BH mass would be $\sim10^{3}\ M_{\odot}$. 
This supports the IMBH interpretation with sub-Eddington accretion rate, although ULXs spectra do not clearly resemble the typical 
low/hard or high/soft state spectra of BH X-ray binaries (BHXBs). 
However, studies of Galactic BHXBs showed that the derived 
temperature from the disk component is reliable only when the spectrum is dominated by the disk emission 
\citep{done2006}. As the contribution from the PL tail component becomes significant 
the derived disk temperature is extremely unreliable. 

ULXs have been observed to show various types of 
spectral shapes. High quality XMM-Newton data revealed spectral curvature associated with the hard X-ray emission component in the 2-10 keV band \citep{stobbart2006,gladstone2009,kajava2012}.
Fits with phenomenological spectral models allowed for the identification of four main 
types of spectral shapes: soft ultraluminous, hard ultraluminous, 
broadened disk and super soft ultraluminous. 
These four types have been ascribed to different accretion regimes
 \citep[for more details see][and references therein]{kaaret2017}. 

\begin{figure*}[]
    \centering
    \includegraphics[width=0.49\textwidth]{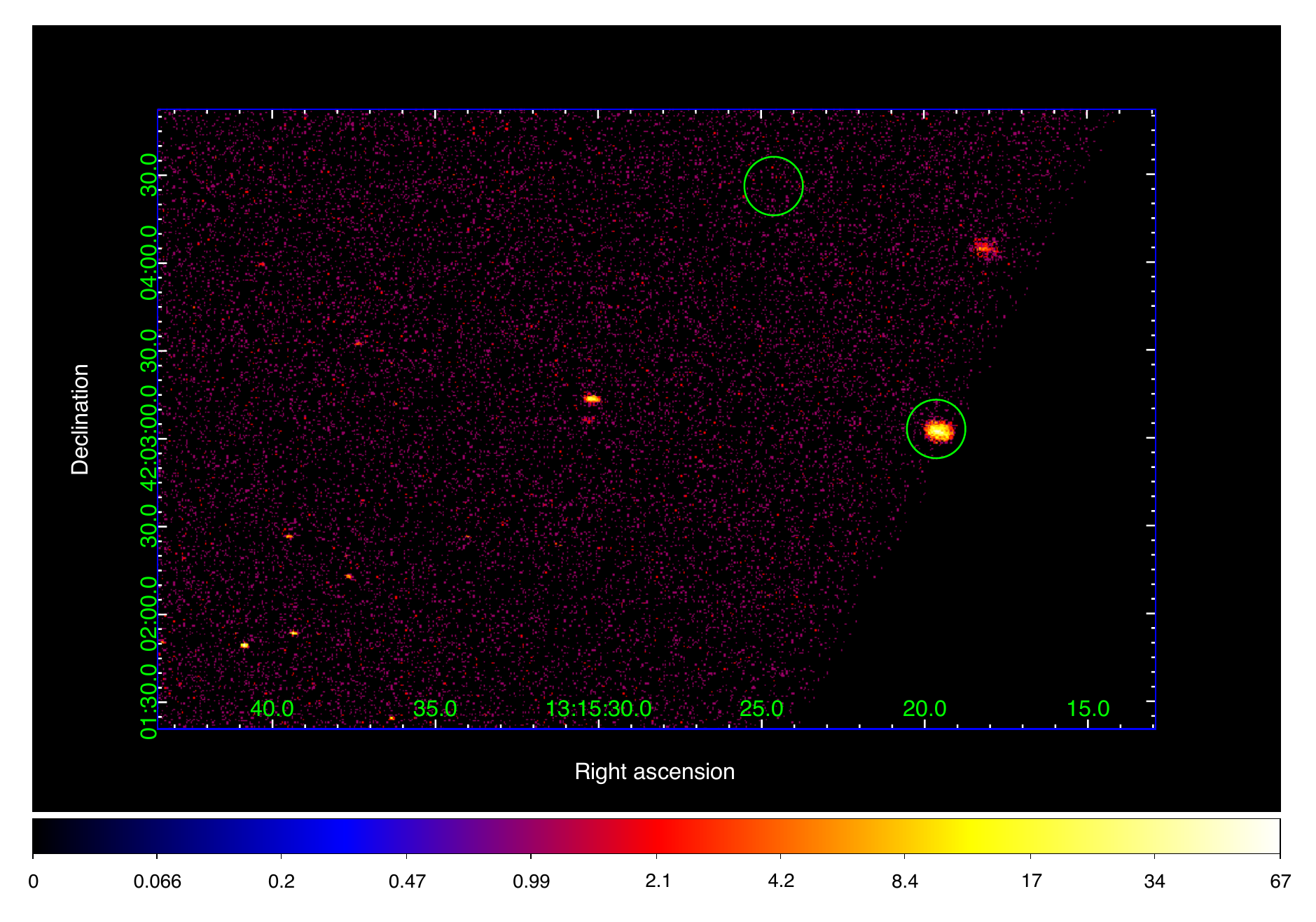}
    \includegraphics[width=0.49\textwidth]{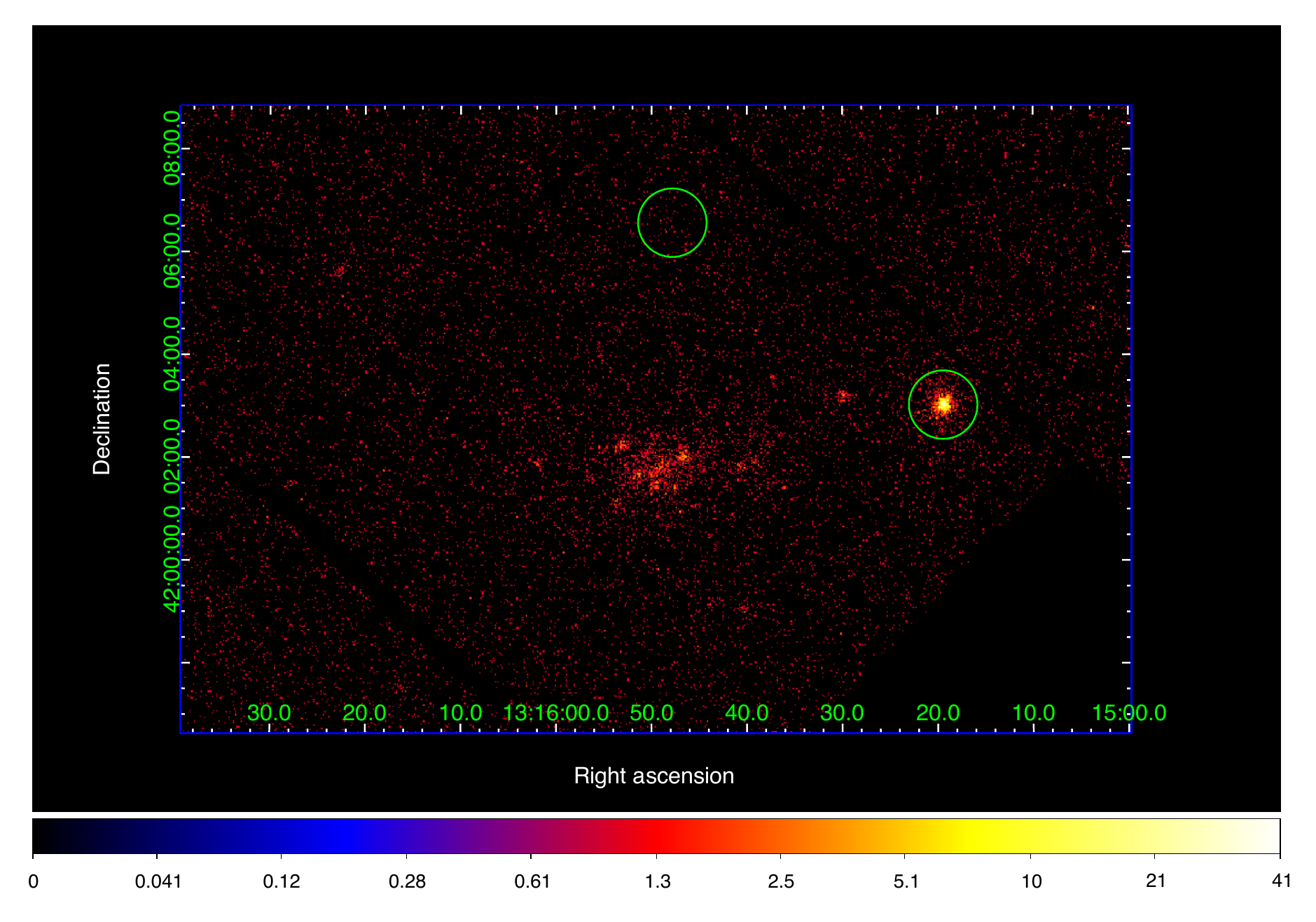}
    \caption{The X-ray  image in counts/pixel  displays the field of view of \emph{Chandra} ACIS-S (left panel) and 
    XMM-Newton EPIC-MOS1 (2007--05--28; right panel) detectors. 
    The green circles containing bright pixels show NGC 5055 X-1, while the empty green circles represent the region from where the 
    background was extracted. \emph{Chandra} spatial resolution allowed us to extract source photons from a circular region of $6''$ radius, while for 
    XMM-Newton we used circular  regions of $40''$. }
    \label{fig:obs}
\end{figure*}

\begin{table}[]
    \centering
    \setlength{\tabcolsep}{6pt}
    \begin{tabular}{cccrr}
        \hline\hline
        \multicolumn{2}{c}{Mission} & & GTI & Net total \\
       Detector & ObsID & Date  & [ks] & Counts \\
        \hline
        \multicolumn{2}{c}{\emph{Chandra}}&   &  &  \\
        ACIS-S & 2197 & 2001.08.27&  28.3 & 2283 \\
        \hline
        \multicolumn{2}{c}{XMM-Newton} & &   & \\
        MOS1 & \multirow{3}{*}{0405080301} & \multirow{3}{*}{2007.05.28}  & 10.9 & 1272  \\
        MOS2 &  &  & 10.9 & 1336  \\
        PN & &  & 8.9 & 4016 \\
        \hline 
       \multicolumn{2}{c}{XMM-Newton} & &  & \\
        MOS1 & \multirow{3}{*}{0405080501} & \multirow{3}{*}{2007.06.19} & 16.1 & 1224 \\
        MOS2 &  &  & 7.8 & 705 \\
        PN & &  & 3.05 & 1149 \\
        \hline
     \end{tabular}
     \caption{Details of X-ray observations analyzed in this paper.
    The detectors used are listed in column 1, observation ID given in column 2,
     date given in column 3.  The exposure good time intervals (GTI), and 
    the net total counts  are listed in column 4 and 5 respectively.
    }
    \label{tab:obs}
\end{table}

\cite{king2001} suggested that most ULXs have a stellar mass compact object accreting at 
super-Eddington rate and their high luminosity results from the beaming of the X-ray emission due to a geometrically thick wind outflowing from 
the disk. Fitting of some ULX spectra with MCD plus PL components, \cite{feng2007} and \cite{kajava2009} found that the temperature of the soft 
component is inversely correlated with luminosity $L\propto T^{-3.5}$, which is at odds with the $L\propto T^4$ 
relation expected from theory and observed in BHXBs \citep{gierlinski2004}. \cite{king2002} 
showed that this inverse relation is actually expected when the source is beamed and accreting at 
super-Eddington rate. Motivated by the observed soft temperature-luminosity relation
\cite{king2009} derived an empirical relation describing the scaling of the beaming factor with the inverse of the square of the accretion rate, 
which implies that the most luminous ULXs are highly collimated.
 However, in the presence of a powerful disk wind one would expect that at large distances 
 from the BH the wind becomes optically thin, and emission and/or
 absorption lines associated with radiation passing through the partially ionized optically thin phase of 
 the wind can be observed. Such features have been found in recent 
studies of NGC 5408 X-1 and NGC 6946 X-1 by \cite{middleton2014}, and of NGC 1313 X-1
and NGC 5408 X-1 by \cite{pinto2016}.

Here, we report on the analysis of \emph{Chandra} and XMM-Newton observations of an extreme ULX in the outskirts of the spiral galaxy NGC 5055 (M63). The source  NGC 5055 X-1, was 
serendipitously discovered by \emph{ROSAT} 
High Resolution Imager (HRI)  \citep{roberts2000}. 
Follow-up observations were carried out by   \emph{Chandra}, XMM-Newton 
and \emph{Swift}.  In spite of being very luminous in the X-ray band, reaching 
$\sim2.3\times10^{40}\ \rm erg\ s^{-1}$ \citep{swartz2011}, the source has received very 
little attention and none of the available X-ray observations are reported in the literature. 

In this paper, we perform the first systematic analysis of X-ray 
observations of NGC 5055 X-1, using three longest \emph{Chandra} and XMM observations.
We carried out X-ray timing and spectral analysis, using phenomenological
 models available in {\sc xspec} fitting package \citep{arnaud1996}.
 The data reduction process is presented in Sec.~\ref{sec:obs}. The timing and spectral analysis 
are presented in Secs.~\ref{sec:time} and \ref{sec:spec}, respectively. The most important correlations between physical parameters are 
shown in Sec.~\ref{sec:results} and the conclusions are reported in Sec.~\ref{sec:conc}.

\begin{figure*}[t]
    \centering
    \includegraphics[trim={1.5cm 3cm 1.5cm 2cm},width=0.33\textwidth, angle=0]{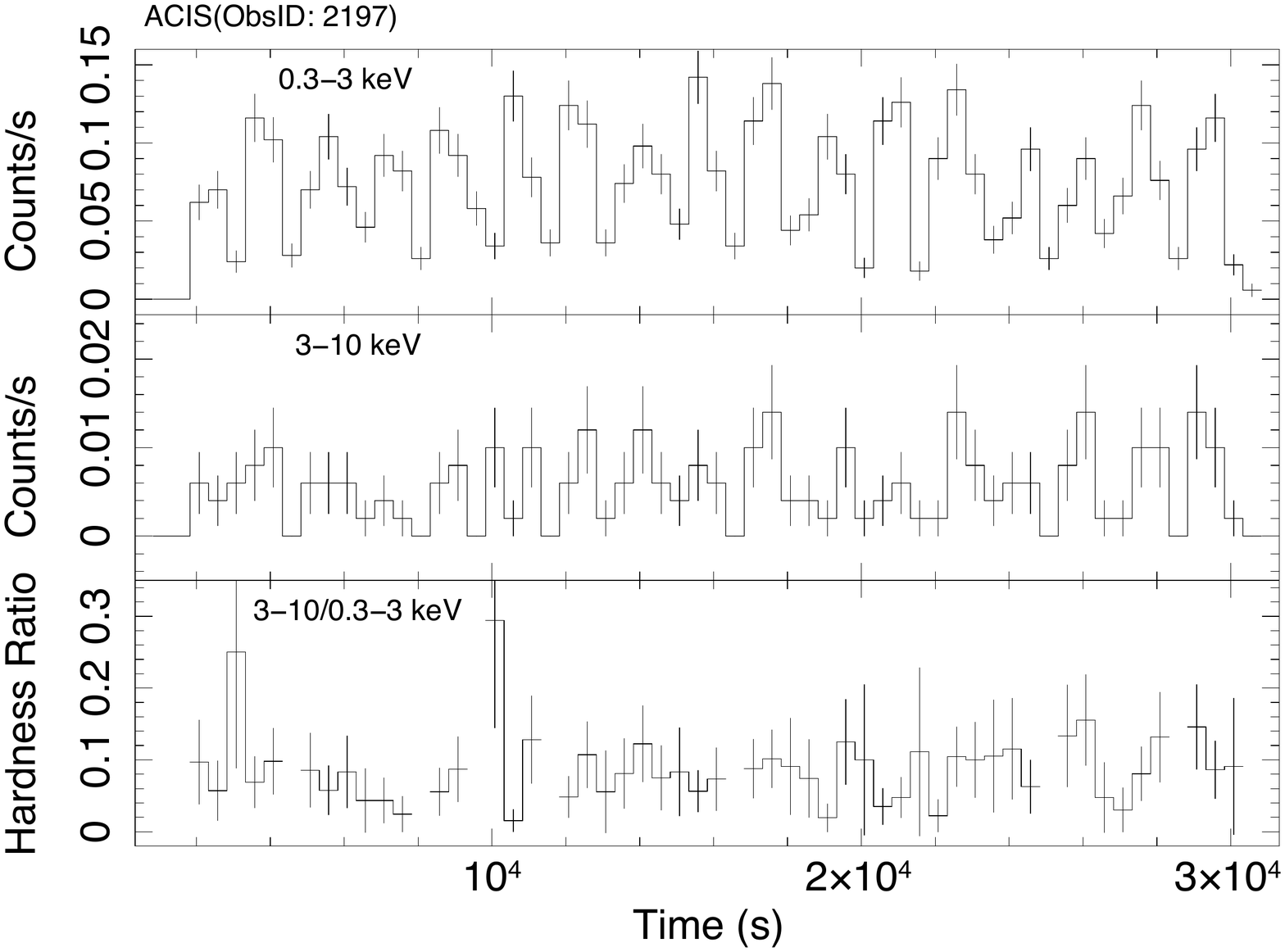}
    \includegraphics[trim={1.5cm 3cm 1.5cm 2cm},width=0.33\textwidth, angle=0]{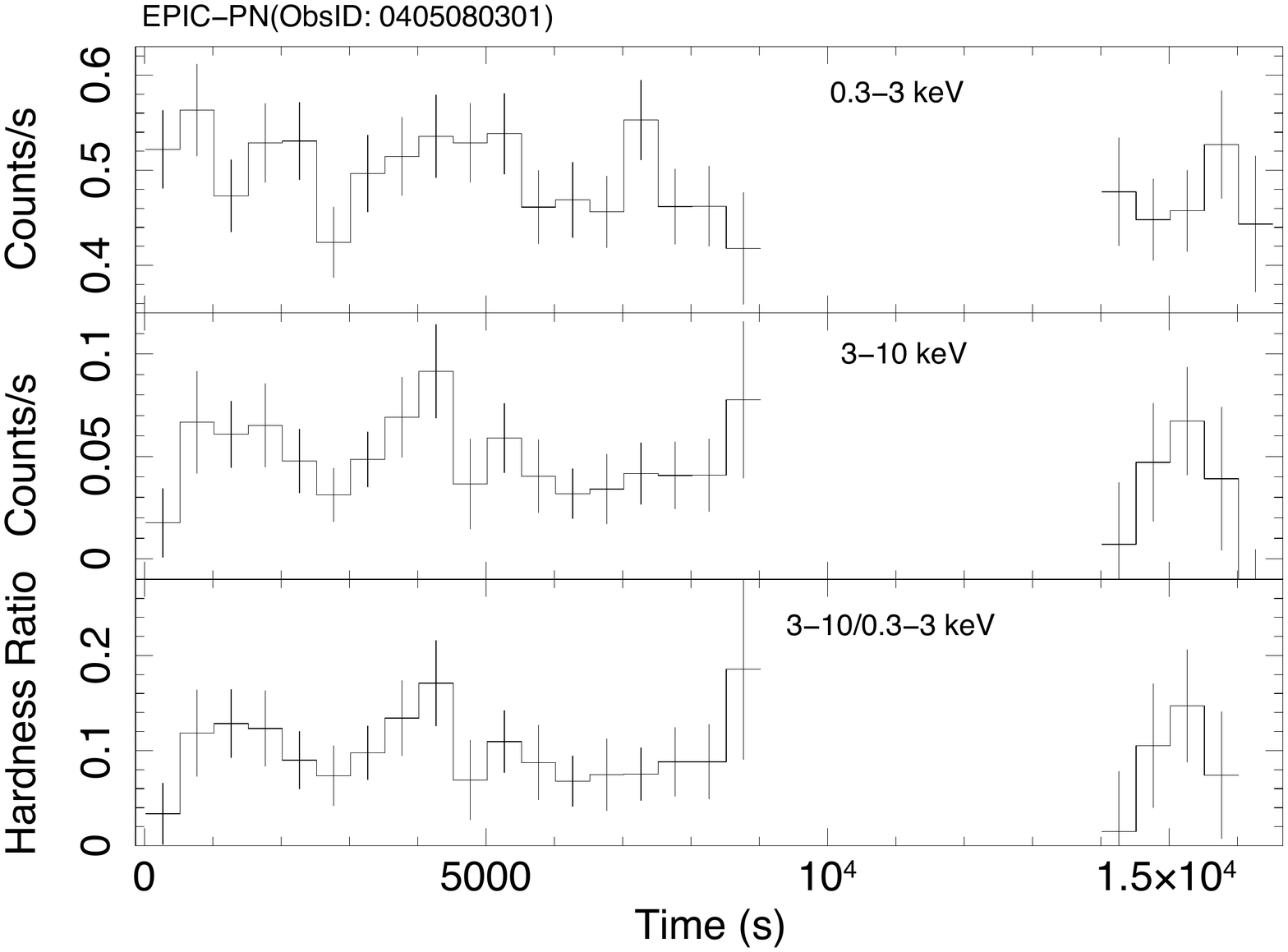}
    \includegraphics[trim={1.5cm 3cm 1.5cm 2cm},width=0.33\textwidth, angle=0]{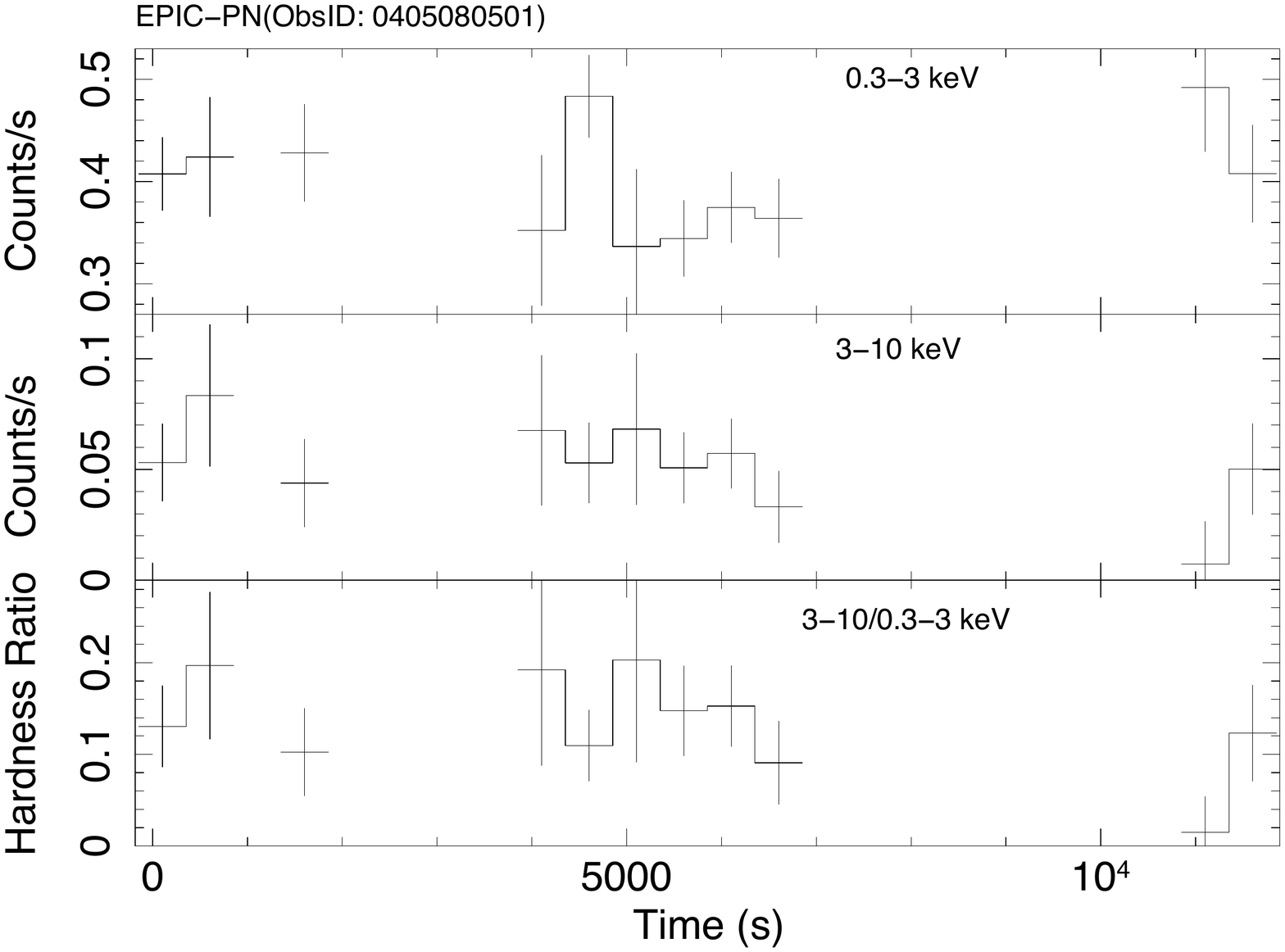}
    \caption{The \emph{Chandra} ACIS-S and XMM-Newton EPIC-PN lightcurves of NGC 5055 X-1 for the three observations at different epochs (the observation IDs are 
    indicated in each panel). Soft 0.3-3 keV and hard 3-10 keV energy band lightcurves are shown in the upper and middle panels 
    respectively, while  - bottom  panels show the hardness ratios for each set of data.
    All the lightcurves are re-binned with bin size 500 s to have higher signal to noise ratio. The zero times in the plots correspond to  2001--08--27  02:13:48, 
    2007--05--28 07:59:14, and 2007--06--19 10:56:29 for panels from left to right, respectively.}
    \label{fig:xmmlight}
\end{figure*}

\begin{figure}[]
    \centering
    \includegraphics[trim={1.5cm 3cm 1.5cm 2cm},width=0.48\textwidth, angle=0]{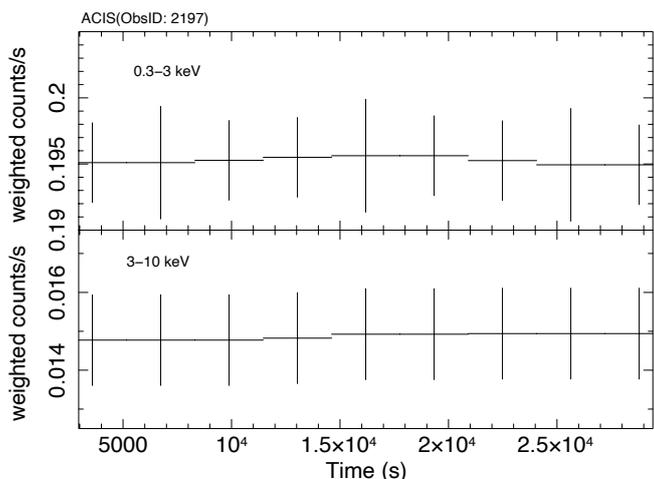}
    \caption{The \emph{Chandra} observation probability-weighted lightcurves, as obtained from the CIAO  \texttt{glvary} tool, i.e. after taking account 
    all instrumental effects.  The zero time corresponds to  2001--08--27  02:13:48. The total probability of variability for the 0.3-3 keV and 3-10 keV 
    lightcurves is 0.02 and 0.07, respectively (see Sec.~\ref{sec:time}).}
    \label{fig:chlight_cor}
\end{figure}

\section{Observations and data reduction}
\label{sec:obs}

The source NGC5055 X-1 is located 
at $\rm R.A.=13^h15^m19.54^s$ and $\rm decl.=42^{o}03^{'}02.3^{''}$. The distance to the galaxy was reported to be 9.2 Mpc 
\citep{tully2013,tikhonov2015, mcquinn2017}, and we use this value throughout this paper. There are many observations of NGC 5055 X-1  
by several X-ray observatories, but most of them are short observations with not enough counts for a meaningful  spectral analysis. Most of the 
\emph{Swift} observations have exposure time $\lsim 10.0~ks$, and total  counts $\sim 160$. The only three long observations available are from  \emph{Chandra} ACIS-S and XMM-Newton EPIC-MOS1, MOS2, and EPIC-PN, with a total of $\sim1000$ net counts, which we used in this study. The details of these observations are given in Tab. \ref{tab:obs}. 

For the data reduction, standard procedures were followed, as described in detail in the following subsections. In all cases NGC5055 X-1
was located in the field of view, but close to the edge of the chip, as shown in 
Fig.~\ref{fig:obs}, where the X-ray images of the source from both satellites are presented. 

\subsection{Chandra data}

We reduced the data from the ACIS-S detector using the standard pipeline Chandra Interactive 
Analysis software CIAOv4.12. NGC 5055 X-1 was detected close to the edge of the field of view (see left panel of Fig.~\ref{fig:obs}), while the 
observatory 
pointed at the center of NGC 5055. We used \texttt{chandra\_repo} to remove the hot pixels, creating new event files 
and producing good time intervals. 
The source and background were extracted from a circular region of $6''$ radius as shown in
Fig.~\ref{fig:obs} left (green circles). The background-subtracted source lightcurve was created using \texttt{dmextract}. The spectrum was generated using 
\texttt{specextract} taking into 
account  the correction 
for the point spread function (PSF) for off-axis sources. 
The auxiliary files and redistribution matrices were generated with \texttt{specextract} using the module \texttt{mkarf}
and \texttt{mkrmf}, respectively.
Given that NGC 5055 X-1 is close to the chip edge (see Fig. 1), the spacecraft dither may 
cause the temporal disappearance of the source from the field of view. We have taken this
into account in our timing analysis
(see Sec.~\ref{sec:time}), and when making auxiliary files for spectral analysis.

\subsection{XMM-Newton}

The data reduction was done using XMM-Newton Science Analysis System (SASv16.0.0) following 
standard procedures. The observation data files (ODF) were processed using 
\texttt{emproc} and \texttt{epproc} 
to create calibrated event lists for EPIC-MOS and EPIC-PN detectors, respectively. A circular region of $40''$ radius 
was chosen to extract source and background counts. We used \texttt{evselect} to generate lightcurves and spectra selecting single and double 
events for EPIC-PN detector and single to quadruple events for EPIC-MOS detector. The source lightcurve 
was corrected from background counts using \texttt{epiclccorr}. 
The auxiliary files and redistribution matrices were generated using \texttt{arfgen} and \texttt{rmfgen}. We note that for XMM-Newton
observation in 2007 June, the EPIC-MOS and EPIC-PN were not observing strictly simultaneously. There is a $\sim52$ minutes of delay between the start 
times of EPIC-MOS and EPIC-PN, which results in a shorter effective (after removing periods of flaring background) exposure time for EPIC-PN, as listed in 
Tab.~\ref{tab:obs}.

\section{Timing analysis}
\label{sec:time} 

The main purpose of our timing analysis is to find out if the source 
shows significant temporal and spectral variability. 
We used \texttt{Stingray}
\footnote{https://stingray.readthedocs.io/en/latest/} \citep{stingray2019} in order 
to construct power spectral density (PSD) from the extracted lightcurves.
 \texttt{Stingray} is an open source spectral-timing Python software package for astrophysical data analysis. 

\begin{table*}[]
    \centering
    \setlength{\tabcolsep}{5pt}
    \renewcommand{\arraystretch}{1.45}
    \begin{tabular}{lrrcccccc}
        \hline\hline
        Data    & $N_{\rm H}$ & $kT_{\rm in}$ & \multirow{2}{*}{$p$} & 
        \multirow{2}{*}{$\Gamma$} & $kT_{\rm e}$ & $F_{(0.3-10){\rm keV}}$ & $L_{(0.3-10){\rm keV}}$ & \multirow{2}{*}{$\chi^2_{\rm red}$} \\
       Model & $10^{20}$~[cm$^{-2}$] & [keV] && & [keV]  &[erg s$^{-1}$ cm$^{-2}$]& [erg s$^{-1}$] & \\
        \hline

        2001.08.27 & &&&&&\\
        \hline
        TBNEW*(MCD+PL) & $15.18^{+1.96}_{-1.73}$ & $0.21^{+0.02}_{-0.04}$ && $2.40^{+0.12}_{-0.11}$ & & $2.30 \times 10^{-12}$ 
        & $2.32 \times 10^{40}$ & 1.22 \\
        TBNEW*(DISKPBB) & $8.30^{+1.86}_{-1.62}$ & $1.27^{+0.25}_{-0.17}$ & $0.50^{+0.03}_{-0.01}$ & & & $1.67 \times 10^{-12}$ & $1.68 \times 10^{40}$ & 1.25 \\
        TBNEW*(MCD+NTHCOMP) & $3.86^{+1.90}_{-1.66}$ & $0.22^{+0.01}_{-0.02}$ && $2.21^{+0.15}_{-0.12}$ & $1.62^{+0.95}_{-0.95}$ & $1.40 \times 10^{-12}$ & $1.41 \times 10^{40}$ & 1.21 \\
        \hline
        2007.05.28 &&&&&\\
        \hline
        TBNEW*(MCD+PL) & $7.76^{+0.88}_{-0.82}$ & $0.24^{+0.01}_{-0.01}$ && $1.99^{+0.08}_{-0.08} $ && $1.55 \times 10^{-12}$ 
        & $ 1.57 \times 10^{40}$  &  1.09 \\
        TBNEW*(DISKPBB) & $5.06^{+0.84}_{-0.77}$ & $1.98^{+0.46}_{-0.30}$ & $0.50^{+0.02}_{-0.01}$ & & & $1.33 \times 10^{-12}$ 
        & $1.34 \times 10^{40}$  & 1.22 \\
        TBNEW*(MCD+NTHCOMP) & $3.72^{+0.87}_{-0.80}$ & $0.25^{+0.004}_{-0.005}$ && $1.85^{+0.09}_{-0.07}$ & $1.78^{+1.56}_{-0.40}$ & $1.28 \times 10^{-12}$ & $1.29 \times 10^{40}$ & 1.08\\
        \hline
        2007.06.19 &&&&&&\\
        \hline
        TBNEW*(MCD+PL) & $7.60^{+1.51}_{-1.34}$ & $0.26^{+0.01}_{-0.01}$ && $1.75^{+0.12}_{-0.11}$ && $1.17 \times 10^{-12}$ 
        & $1.18 \times 10^{40}$ & 1.12 \\
        TBNEW*(DISKPBB) & $5.98^{+1.42}_{-1.26}$ & $2.86^{+0.67}_{-0.67}$ & $0.50^{+0.03}_{-0.01}$ &&& $1.04 \times 10^{-12}$ 
        & $1.05 \times 10^{40}$ & 1.22 \\
        TBNEW*(MCD+NTHCOMP) & $5.53^{+1.51}_{-1.34}$ & $0.27^{+0.006}_{-0.006}$ && $1.59^{+0.10}_{-0.07}$ & $1.62^{+0.60}_{-0.28}$ & $1.02 \times 10^{-12}$ & $1.03 \times 10^{40}$ & 1.12\\
        \hline
    \end{tabular}
     \caption{Best fit parameters obtained from the fits of each set of data. We tied the disk inner radius temperature to the temperature of the soft seed 
     photons for thermalized Compton emission in MCD+NTHCOMP model. For XMM-Newton observations MOS1, MOS2 and PN data were fitted simultaneously. 
     The unabsorbed flux and luminosity in the range 0.3-10 keV estimated from the best-fit model are given in column 7 and 8, assuming a distance 
     to the source of 9.2 Mpc.}
    \label{tab:spec}
\end{table*}

We used the XMM-Newton EPIC-PN lightcurves for timing analysis as the EPIC-PN detector has
 higher full 
frame time resolution (73.4~ms) and effective area than the EPIC-MOS cameras. The lightcurves were extracted 
with a time bin of 0.22~s (3 times the temporal resolution) to have 
enough counts in each bin. 
We used 9.6~s time bin (\emph{Chandra} temporal resolution is 3.2~s) to extract the 
lightcurves from \emph{Chandra} ACIS-S
observation. Fig.~\ref{fig:xmmlight} shows the ACIS-S and EPIC-PN lightcurves 
for the two bands 0.3-3~keV and 3-10~keV, re-binned with bin size 500 s. We do not show 
lightcurves from both MOS cameras in the figure due to their lower S/N ratios. 
The source does not display either flux nor spectral variability. 

Nonetheless, ACIS-S lightcurves show enhanced variability with an apparently quasi-periodic pattern. The hardness ratios are consistent with being 
constant (left most panel of Fig.~\ref{fig:xmmlight}). We constructed PSD from ACIS-S lightcurves to explore the nature of the periodic pattern. To 
this aim, we used the unrebinned lightcurves, with a time resolution of 9.6 s. Each individual lightcurve was divided into 4 segments and the PSD was computed in each 
segment separately. Then,  we averaged the PSD from the 4 segments.

In  \emph{Chandra}
data we found periodic variations corresponding to two peaks 
at 1.4~mHz and 2.8~mHz. Such features are clearly not observed during XMM observations.
While the hypothesis of a quasi-periodic feature in \emph{Chandra} lightcurves disappearing in later XMM observations is tantalizing, we verified this is most likely an instrumental artifact. Indeed, as pointed out in Sect. 2.1, the satellite 
dithering motion combined with the position of the source near the edge chip, may cause the source to periodically disappear. To verify this, we 
used the CIAO tool \texttt{glvary} to search for significant variability in \emph{Chandra} lightcurves. The \texttt{glvary} tool utilizes 
information from the \texttt{dither\_region} tool to correct the instrumental effects. The \texttt{dither\_region} tool calculates the fractional 
area of the source region as a function of time that takes into account for the 
corrections for chip edges, bad pixels, and bad columns. 
As an output, \texttt{glvary} computes the probability that the lightcurve is variable using Gregory-Loredo algorithm \citep{gregory1992}. 
The resulting probability values for the 0.3-3 keV and 3-10 keV lightcurves are 0.02 and 0.07, implying that the source is not significantly variable. In 
Fig.~\ref{fig:chlight_cor}, the \emph{Chandra} probability-weighted lightcurves  
obtained with \texttt{glvary} are presented. We conclude that the source is not significantly variable and the quasi-periodic feature is an instrumental artifact. A more detailed description on how the dither motion 
affects the observed lightcurves for sources located near the chip edge is given in \citet[][]{roberts2004}.

\section{Spectral analysis}
\label{sec:spec}

\begin{figure}
    \centering
    \includegraphics[trim={0 0 0 0},width=0.45\textwidth]{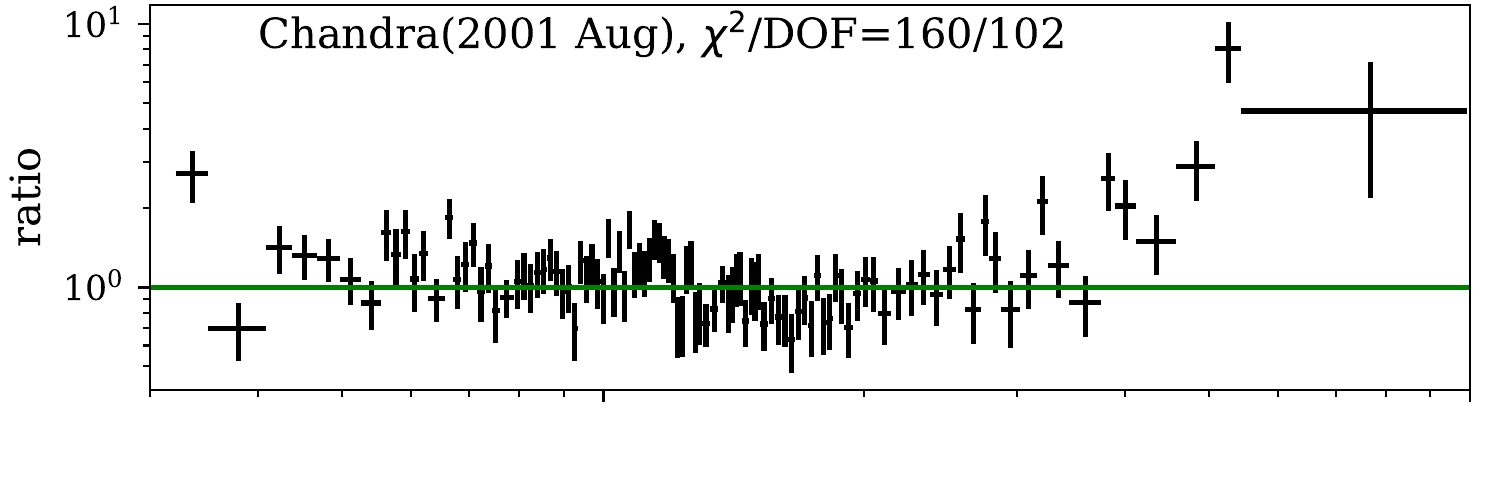}
    \includegraphics[trim={0 0 0 1cm},width=0.45\textwidth]{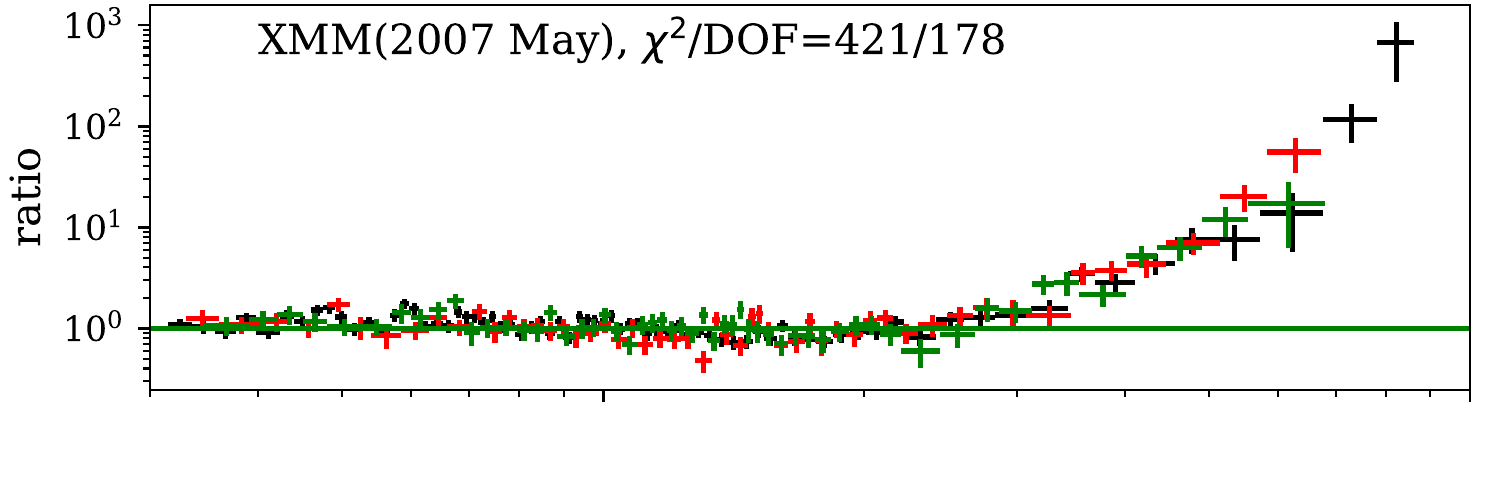}
    \includegraphics[trim={0 0 0 1cm},width=0.45\textwidth]{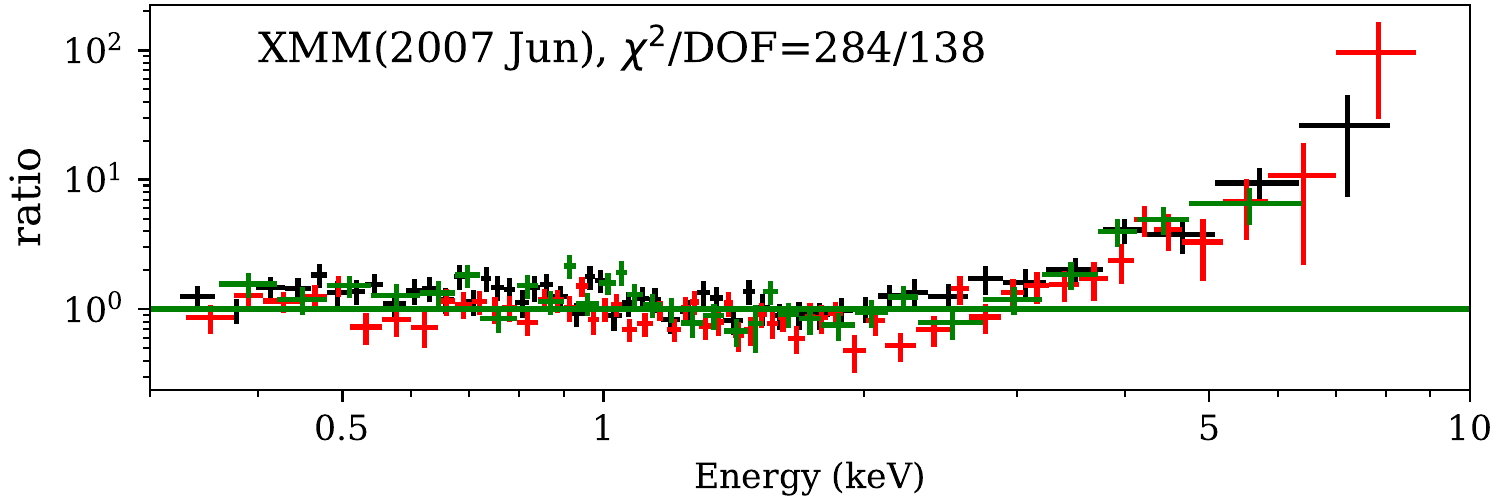}
    \caption{The ratio of data to folded model obtained from the fit of a MCD model, using DISKBB in {\sc xspec}. The MCD model clearly leaves an excess at high 
    energies. In XMM-Newton panels the black, red and green data points are, respectively, the EPIC-PN, EPIC-MOS1 and EPIC-MOS2 spectra.}
    \label{fig:diskbb}
\end{figure}

\begin{figure*}[]
    \centering
    \includegraphics[trim={0 0 0 0},width=0.33\textwidth]{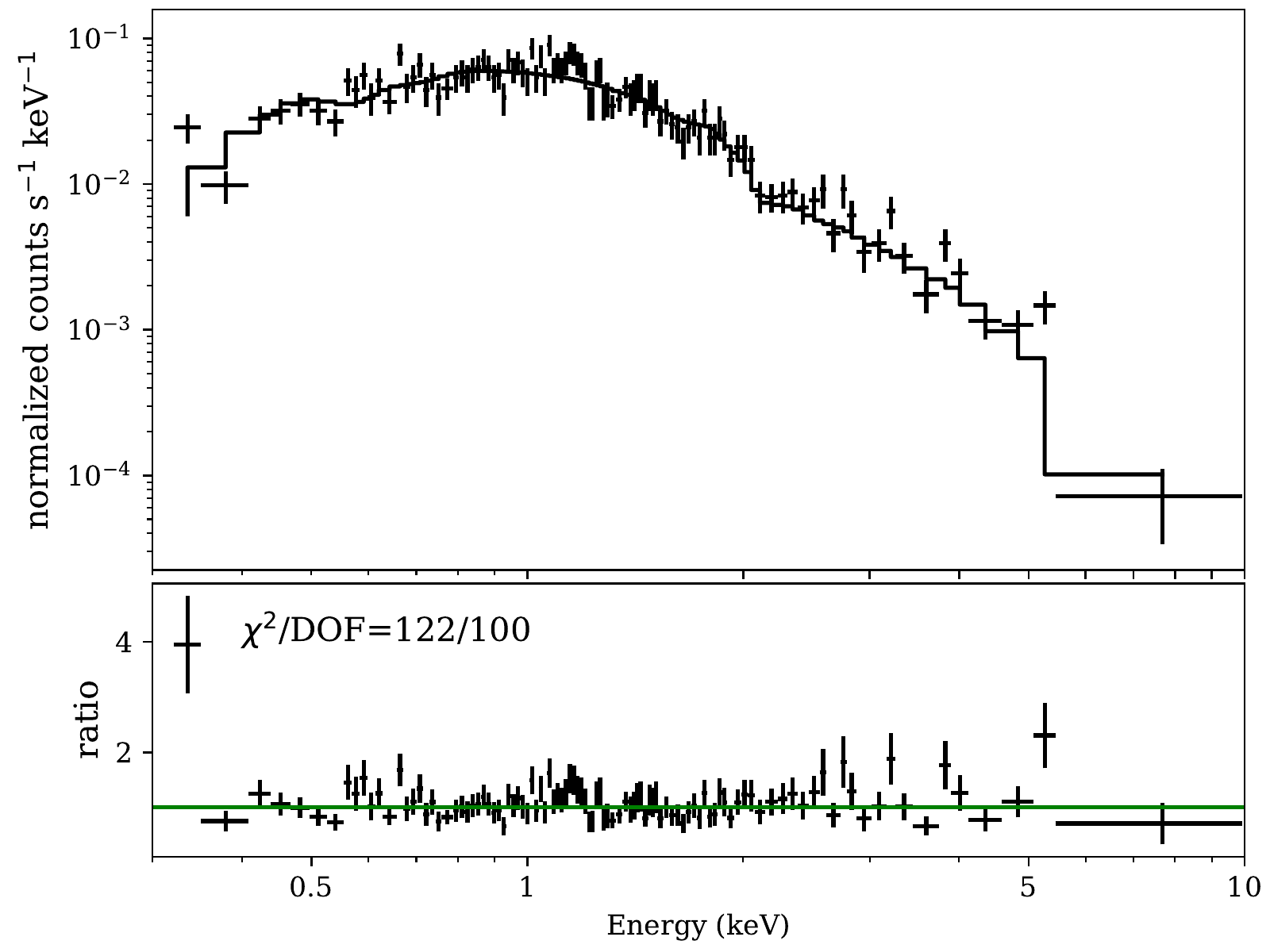}
    \includegraphics[trim={0 0 0 0},width=0.33\textwidth]{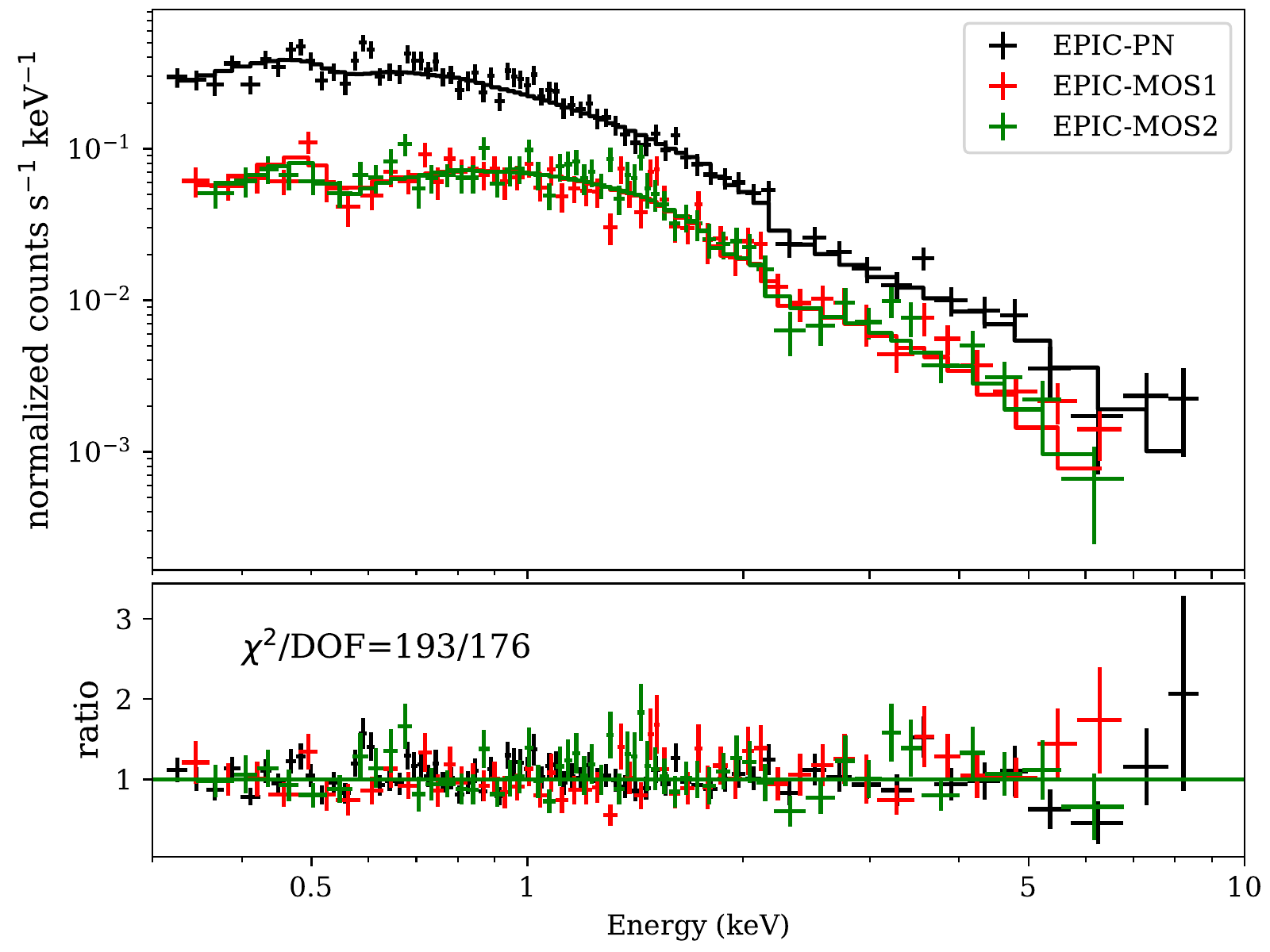}
    \includegraphics[trim={0 0 0 0},width=0.33\textwidth]{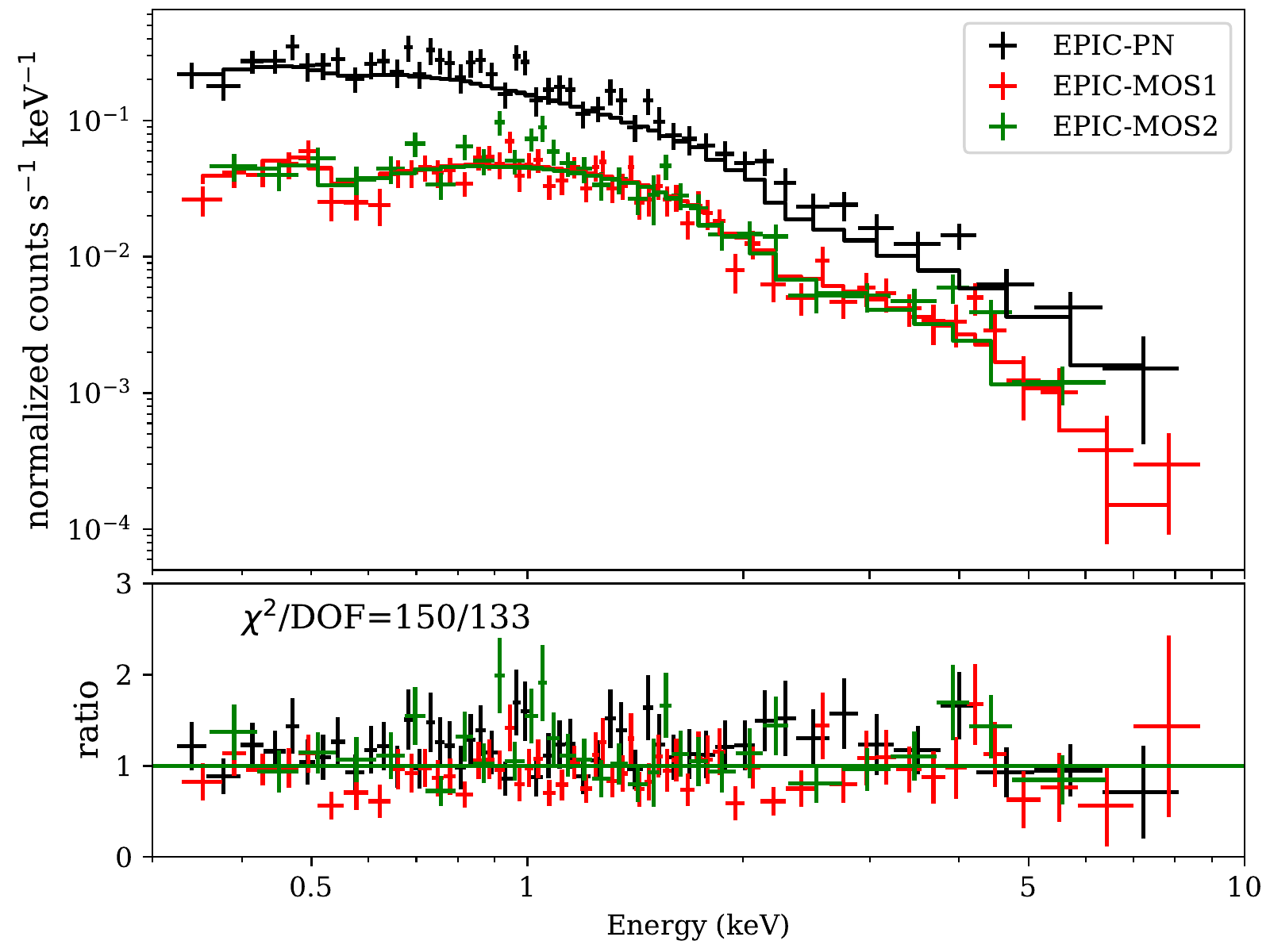}
    \includegraphics[trim={0 0 0 0},width=0.33\textwidth]{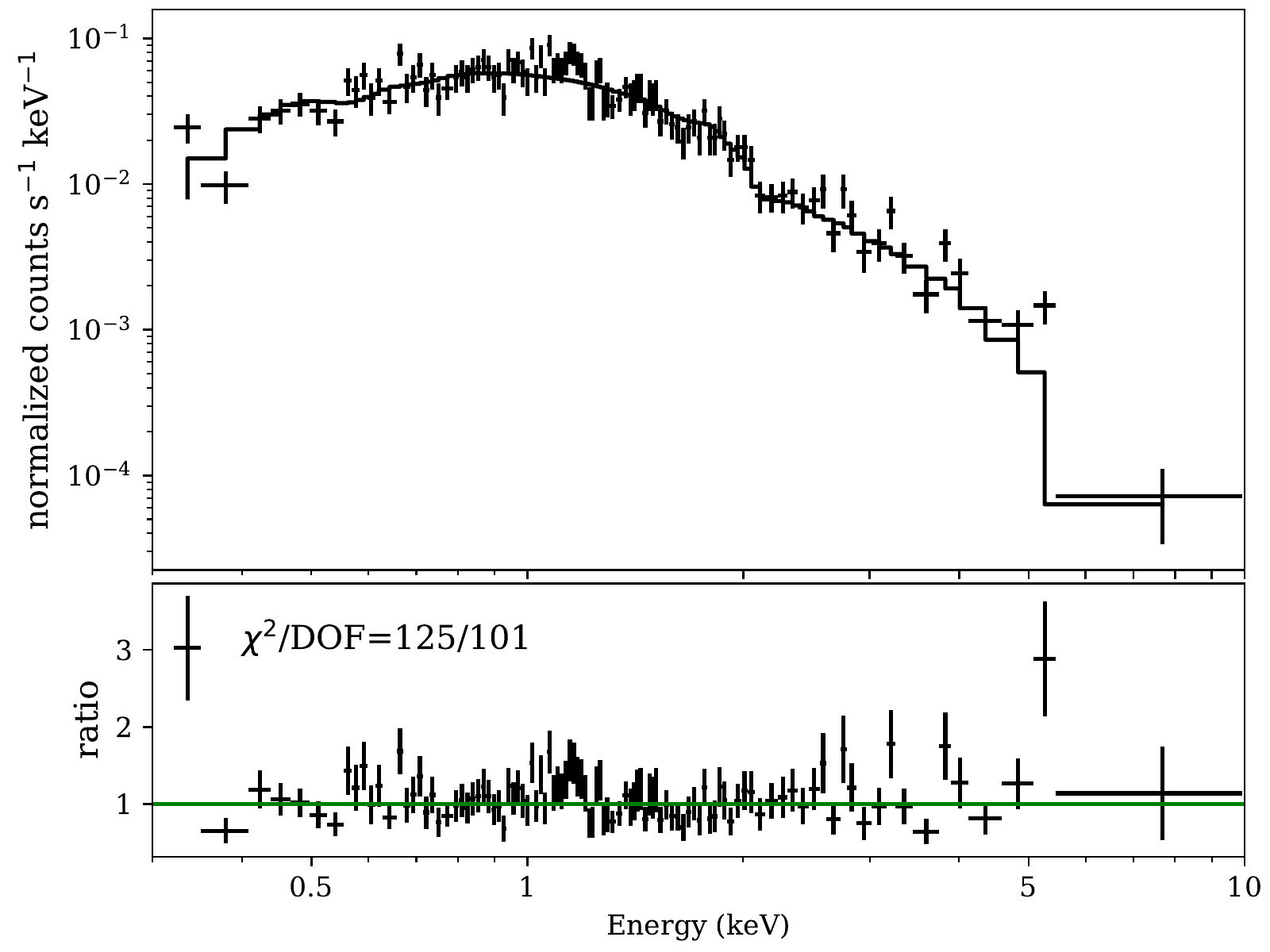}
    \includegraphics[trim={0 0 0 0},width=0.33\textwidth]{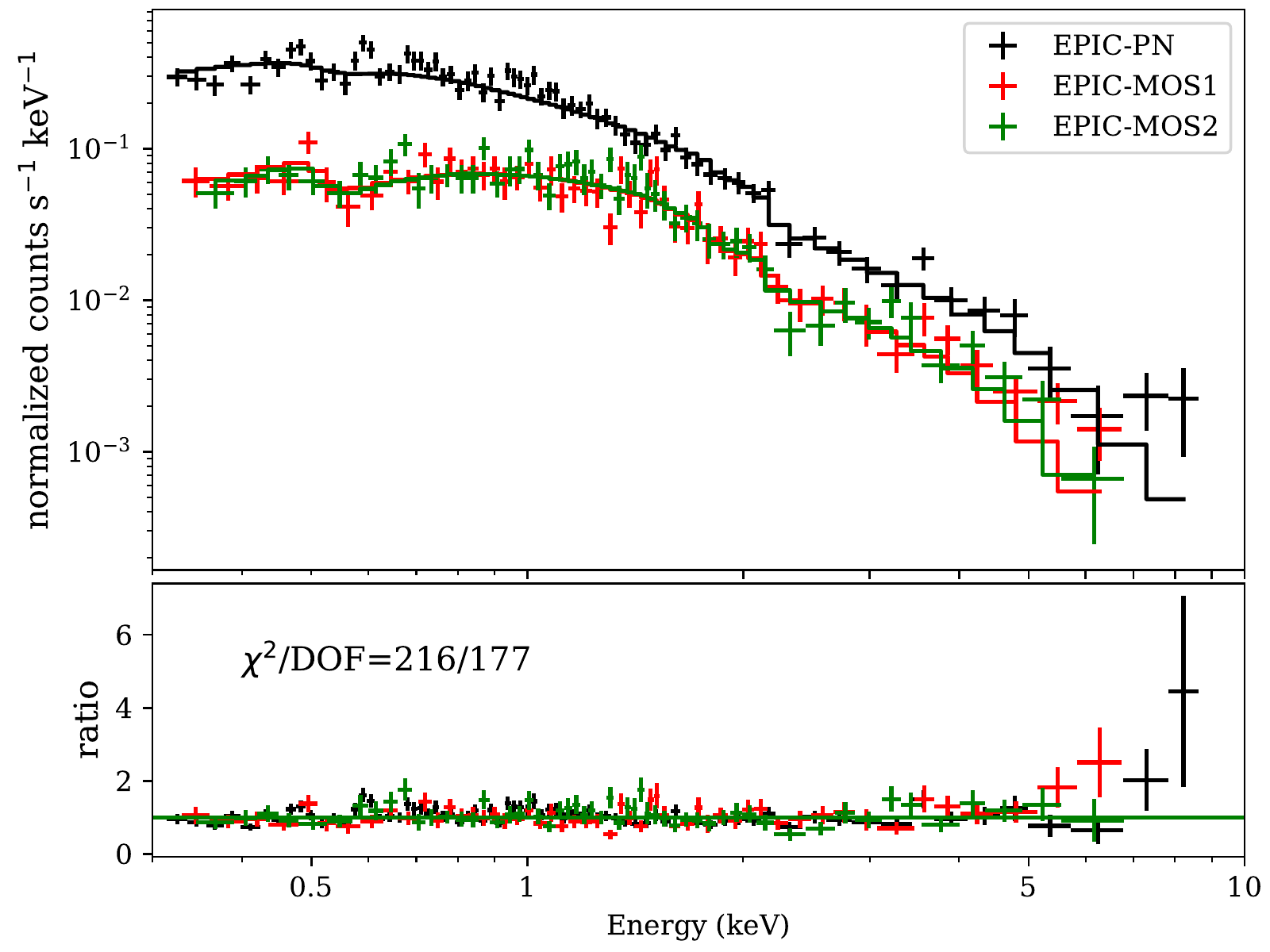}
    \includegraphics[trim={0 0 0 0},width=0.33\textwidth]{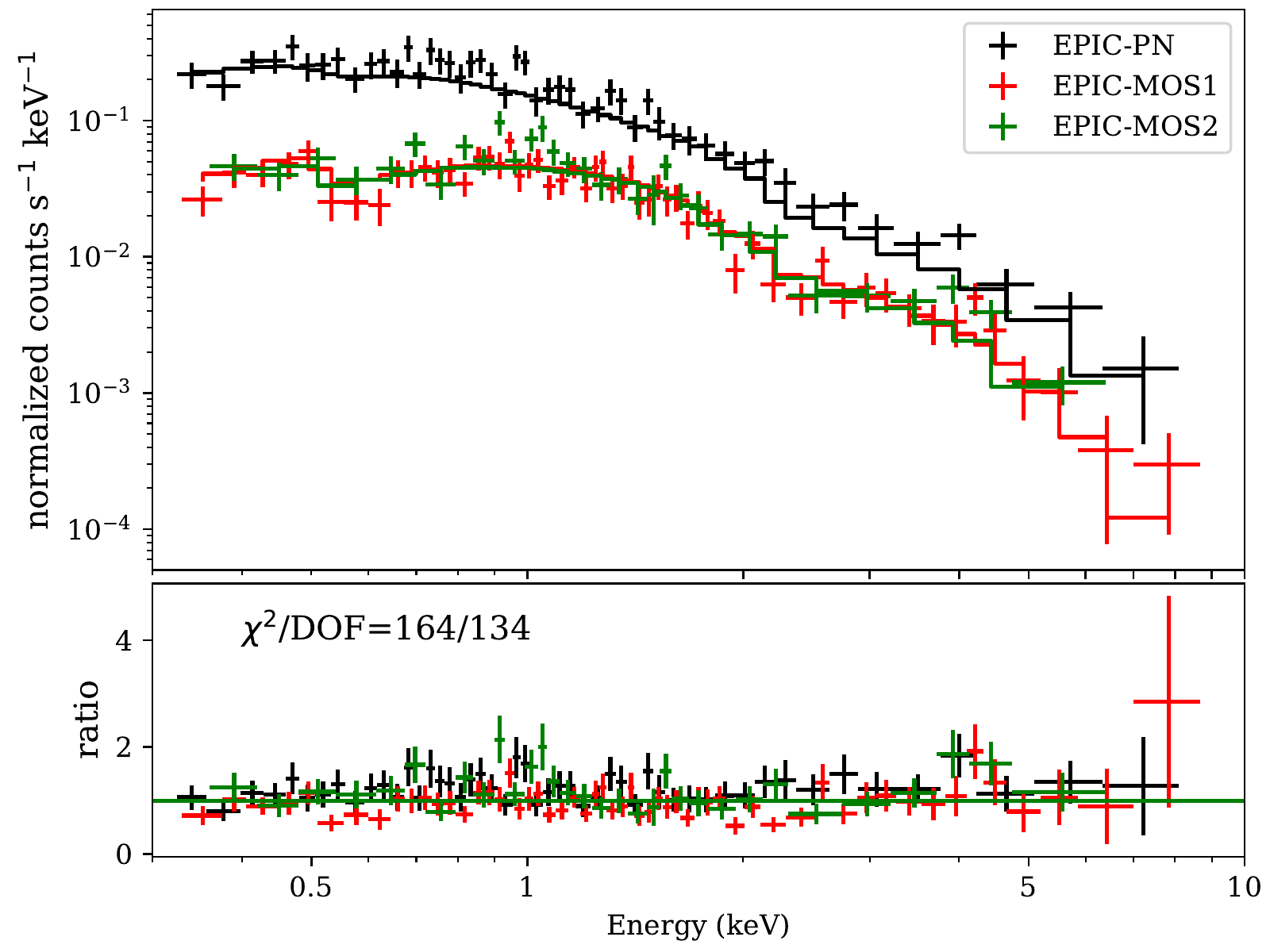}
    \includegraphics[trim={0 0 0 0},width=0.33\textwidth]{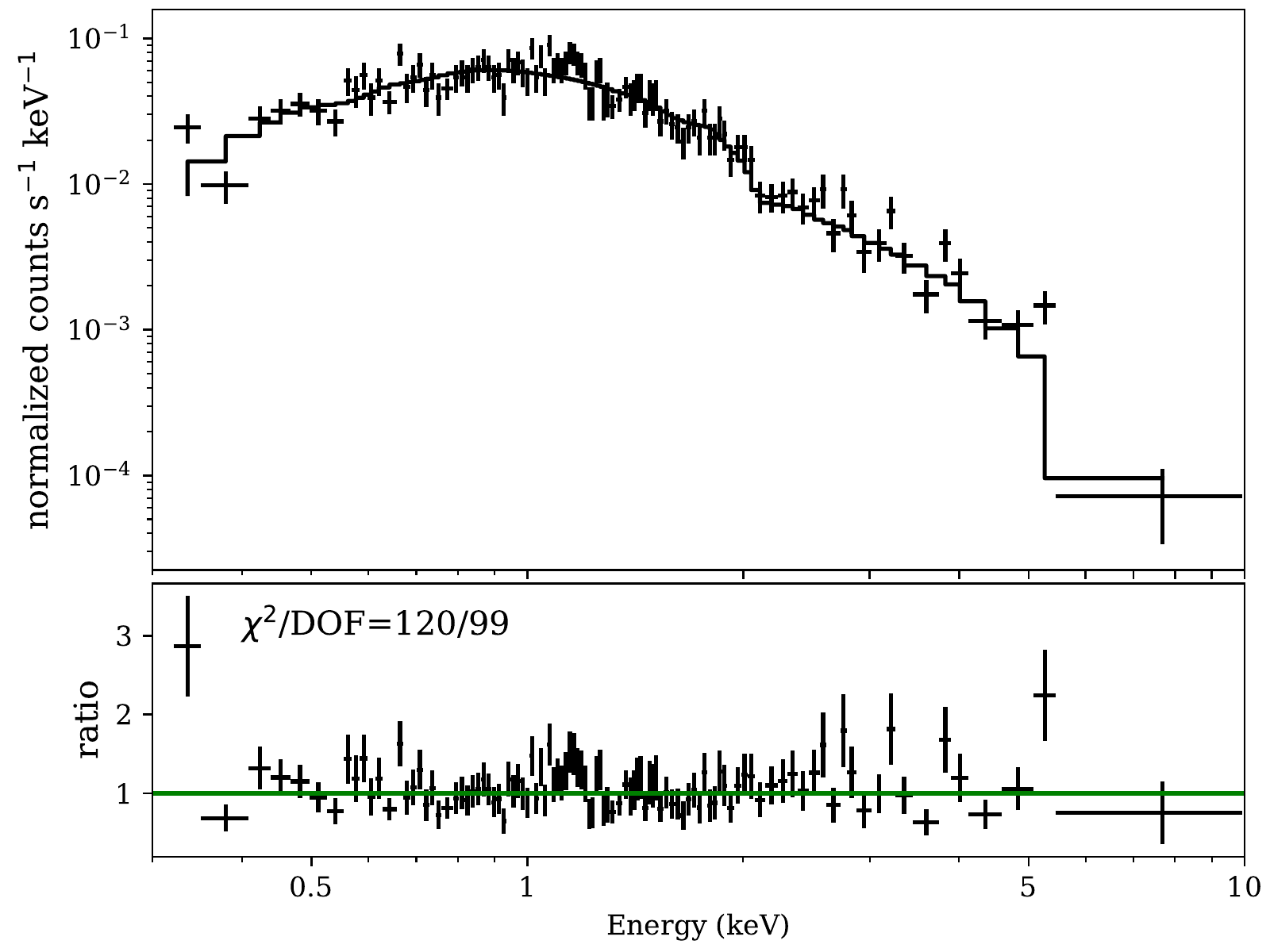}
    \includegraphics[trim={0 0 0 0},width=0.33\textwidth]{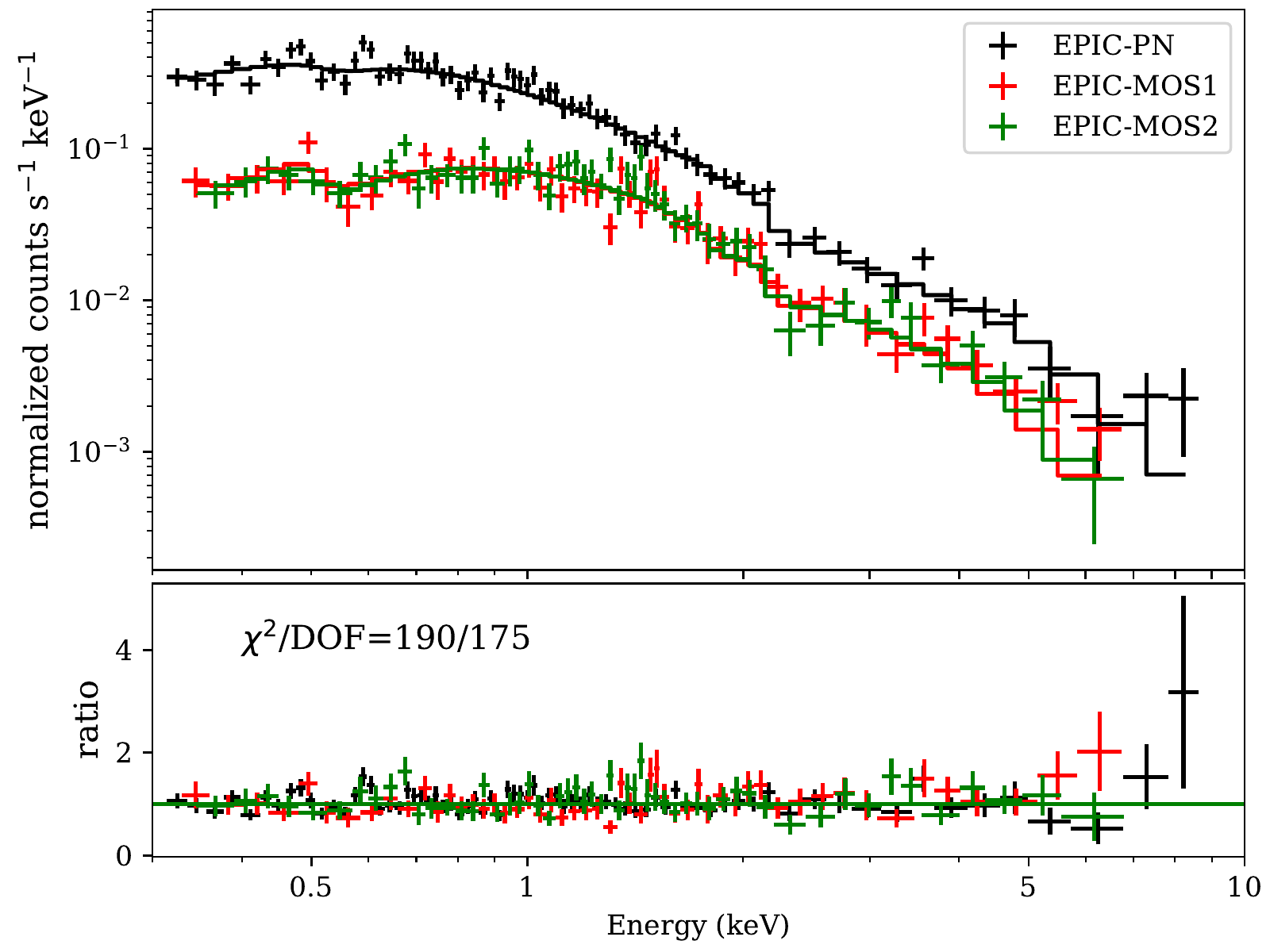}
    \includegraphics[trim={0 0 0 0},width=0.33\textwidth]{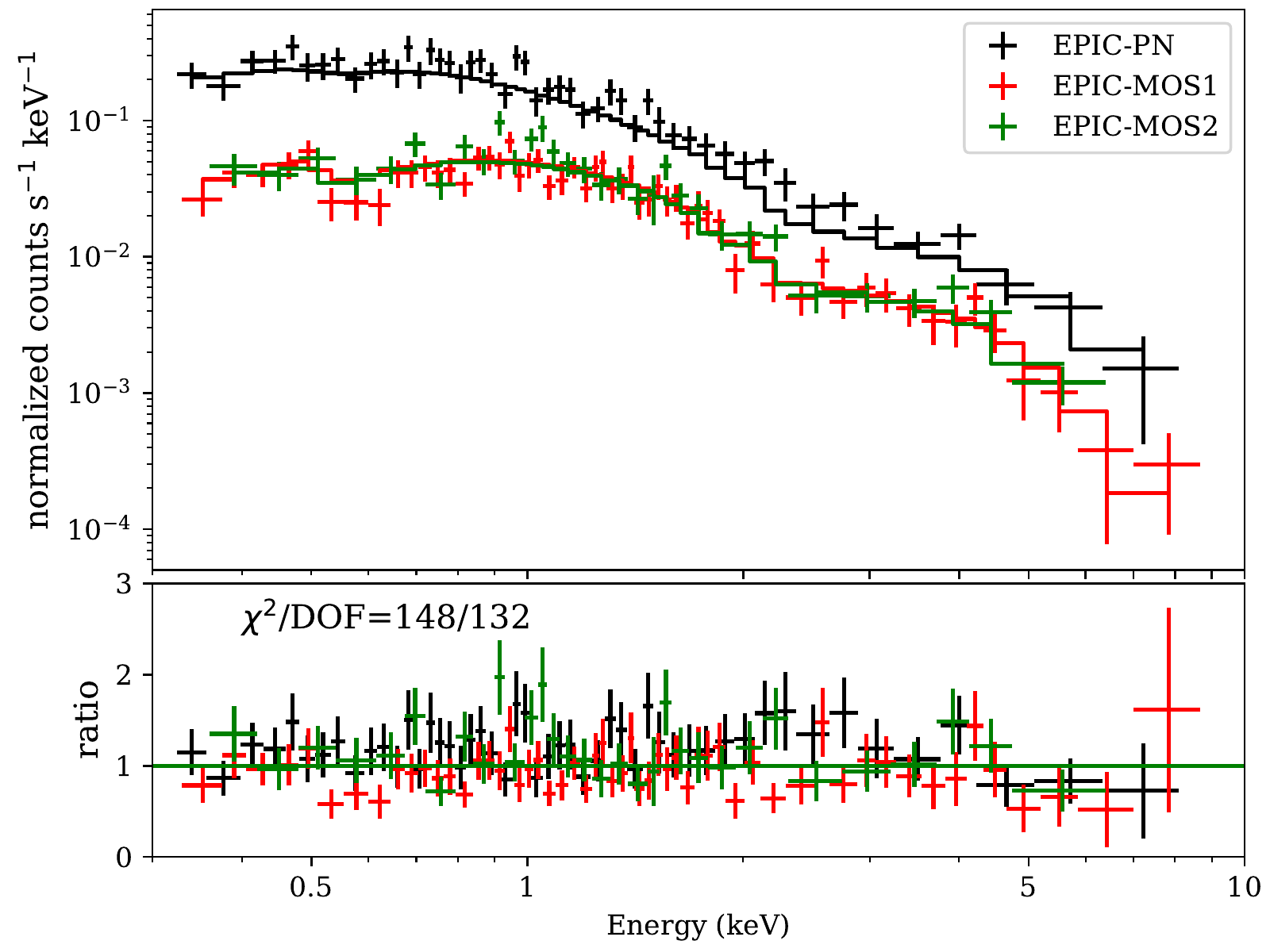}
    \caption{Each subplot shows the data with the folded model in the upper panel, and the data-to-model ratio in the lower panel. In the first row, 
    the results from the fit of the MCD+PL model are presented, in the second row - DISKPBB model, and finally in the third row - MCD+NTHCOMP. The 
    different columns correspond to \emph{Chandra} data - left column,  XMM-Newton 2007--05--28 data- middle column, and 2007--06--19 data - right 
    column.}
    \label{fig:spec}
\end{figure*}

\begin{figure}
    \centering
    \includegraphics[width=0.48\textwidth]{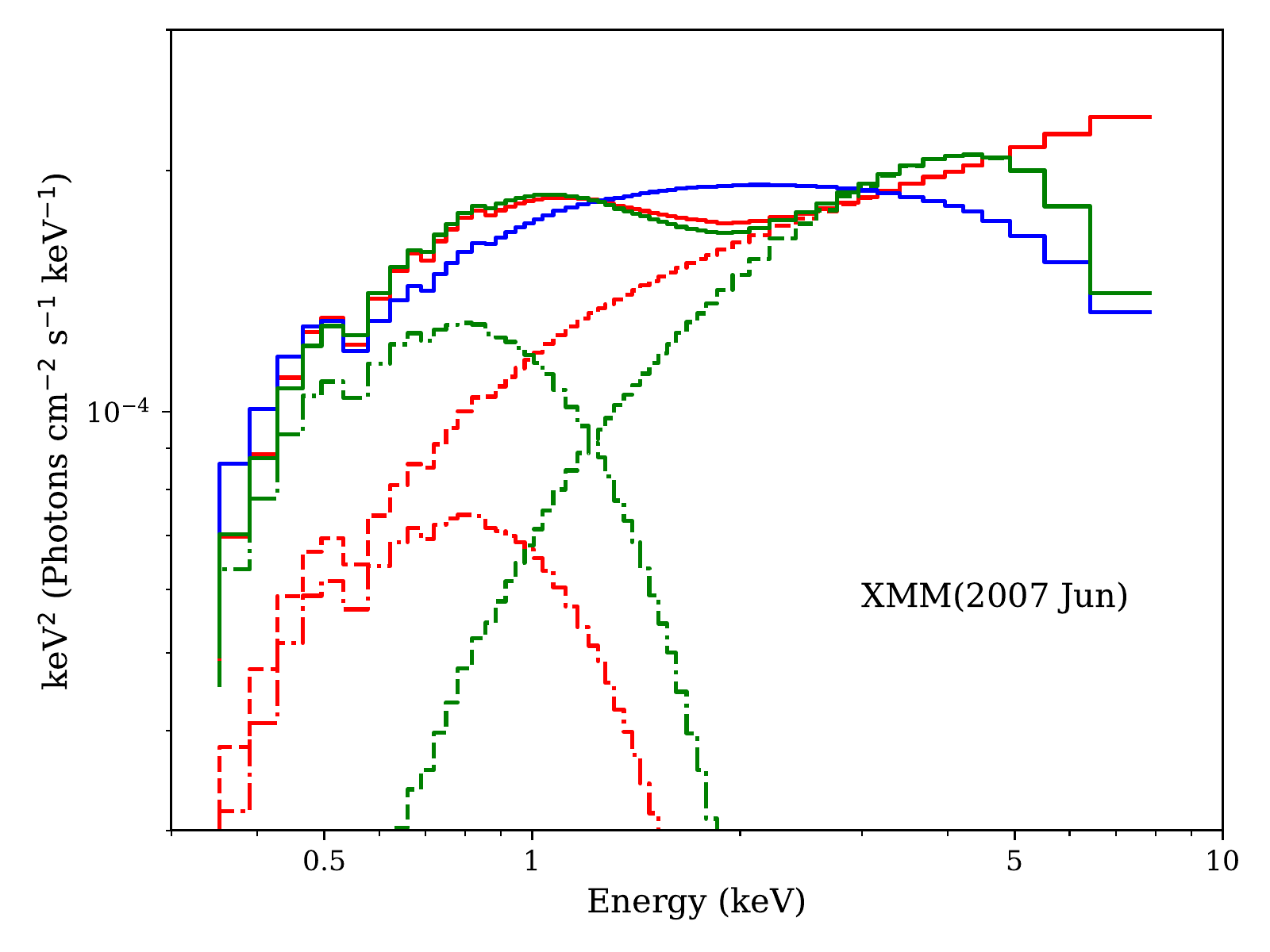}
    \caption{Unfolded absorbed models (solid lines) and model components (dot-dashed line for MCD and dashed line for NTHCOMP and PL) fitted to the 
    XMM-Newton 2007--06--19 data. Red lines represent the MCD+PL model, the blue line is the DISKPBB model, and green lines show the MCD+NTHCOMP 
    model in which the MCD component is the most prominent. 
    }
    \label{fig:model}
\end{figure}    

As the hardness ratios presented in Figs.~\ref{fig:xmmlight} and 
\ref{fig:chlight_cor} (lowest panels) do not show significant indications of spectral variability, we carried out fits to the time-averaged spectra of 
the single observations. 
We tested simple accretion disk models in order to understand 
the basic properties of the source. We utilized the {\sc xspec}v12.10.1 \citep{arnaud1996} software for spectral analysis and different spectral 
models were used to fit the data: we first fit a simple MCD using the DISKBB model in {\sc xspec}, then we added a POWERLAW component (MCD+PL); next we 
considered slim disk emission \citep[DISKPBB  model,][]{mineshige1994,hirano1995,
watarai2000,kubota2004,kubota2005}; and finally we considered a thermal Comptonization component 
due to a hot corona \citep[MCD+NTHCOMP,][]{zdziarski96,zycki99}.
The spectra are fitted in the band 0.3-10 keV and re-binned to have minimum of 20 counts in each energy bin.
The effect of interstellar absorption was accounted for using 
TBNEW\footnote{https://pulsar.sternwarte.uni-erlangen.de/wilms/research/tbabs/}. We let the hydrogen column density $N_{\rm H}$
free to vary, and keeping in mind that the value of Galactic absorption towards NGC 5055 is estimated to be $\sim3.57\times10^{20}\ 
\rm cm^{-2}$ \citep{bekhti2016}. 

Following standard recommendations (see SAS website 
\footnote{https://www.cosmos.esa.int/web/xmm-newton/sas-thread-epic-merging}),
we did not co-add the spectra from each XMM-Newton detector, but we fit them simultaneously.
We used the same models  for \emph{Chandra} and  XMM-Newton data sets, and as a result we obtained constraints on physical parameters from three 
different epochs. 

As a first step, we tested the well known MCD model, representing emission from 
a standard \cite{SS73} thin disk. The standard disk model is appropriate 
when the source accretion rate is below the Eddington value, assuming that  
the accretion disk is geometrically thin ($h/r<<1.0$) and the radiation is emitted locally 
as black body from optically thick gas.  
When we fit each data set with a single MCD,  the reduced $\chi^2$ is high, 
and the ratio of data to the model significantly differs from unity as shown in 
Fig.~\ref{fig:diskbb}.  Furthermore, the best-fit $N_{\rm H}$ value is very low, 
much lower than the estimated value for Galactic absorption, which suggests that the fit is unphysical. 

Adding an extra PL component to the MCD model, we obtained an excellent 
fit for all the three data sets, as described in Tab~\ref{tab:spec}
and shown in Fig.~\ref{fig:spec} (upper panels).  The inner accretion disk temperature is relatively 
low compared to the values typically observed in Galactic BHXBs in the soft state \citep{gierlinski2004}. Assuming the source is in a state similar to the 
soft state, then the low temperature is consistent with an IMBH of $\sim 10^3 M_{\odot}$. 
The first  XMM-Newton data set (2007--05--28) has higher flux 
than the second data set (2007--06--19), but slightly lower than in \emph{Chandra} data. 
The required Galactic $N_{\rm H}$ in \emph{Chandra} 
is twice as large than in both XMM-Newton data sets. 

However, the assumption of a geometrically thin and optically thick disk is not valid when the source is close or above the Eddington limit. 
The luminosity of NGC 5055 X-1 is very high. If the accretor is a stellar mass BH, then the accretion rate is well above 
the Eddington rate.  Assuming a 10 $M_{\odot}$ BH and isotropic emission
we infer $\dot M_{\rm acc}>10^4\ \dot M_{\rm Edd}$, and bolometric luminosity as:
\begin{equation}
L_{\rm bol}=L_{\rm Edd}[1+\text{ln}(\dot M_{\rm acc}/\dot M_{\rm Edd})].
\end{equation} 
In the regime of super-Eddington accretion, the disk becomes geometrically 
thick ($h/r\sim1$) and the photon diffusion timescale in the vertical direction becomes much longer than the radial in-fall timescale. 
This allows some of the photons to be advected radially into the BH rather than emitted locally which leads to a flatter 
effective temperature profile \citep{abramowicz88,watarai2000}. The high luminosity of NGC 5055 X-1 suggests that the accretion rate may be 
extremely high. In this case one would expect a significantly different temperature profile than in the standard thin disk model. The DISKPBB 
slim disk model allows the disk radial temperature profile to be fit to the data, with the local disk temperature  $T(r) \propto r^{-p}$, 
where $p$ is a free parameter. The standard MCD model is recovered if $p=0.75$ and has been usually 
used in previous studies of ULXs.

The DISKPBB model while providing an overall acceptable description of the data, yields slightly worse fits, as shown in Fig.~\ref{fig:spec} middle 
row panels. All best-fit parameters 
are described in Tab~\ref{tab:spec}. The Galactic $N_{\rm H}$ is lower than in the case of 
MCD+PL model giving higher inner disk temperature, i.e. $kT_{\rm in} = 1.27$, 1.98 and 2.86 keV for the first, second and third data set, respectively.
 The temperature profile obtained from this model is the same in all data sets i.e. 
$T(r)\propto r^{-0.5}$ which is significantly different from the thin disk model ($p=0.75$).

The fit  with MCD+NTHCOMP 
model presented in Fig.~\ref{fig:spec} bottom panels, gives the same reduced $\chi^2$ 
and the same disk $T_{\rm in}$ as the MCD+PL model (see Tab.~\ref{tab:spec} for details). 
The only difference is the value of Galactic absorption, which in the case of the MCD+NTHCOMP model has lower values than in other models, but
in agreement with independent measurements of  $N_{\rm H}$  \citep[as reported by][]{bekhti2016},
for the first two data sets. 
This means that the normalization, 
and hence the contribution, of the MCD component
is higher when fitted together with the NTHCOMP component than with a simple PL component, 
as illustrated in Fig.~\ref{fig:model}. Even if the fit statistic does not allow us to 
differentiate between those two models, the inferred value of Galactic absorption 
slightly favors the MCD+NTHCOMP model. The second reason to favor the model with thermal Comptonization is that in this model the two components are physically linked in a self-consistent way, i.e. 
the temperature of the corona and hence the low energy cut-off of 
NTHCOMP depends on 
the inner disk temperature. On the contrary, the PL component does not have a low energy cut-off, and therefore the model is less physical.  
The temperature of the corona from the MCD+NTHCOMP best-fit model is quite 
low $kT_{\rm e} \sim 1.68-1.78$ keV. This is commonly observed in ULXs \citep{gladstone2009} and can be ascribed to a cool corona, reminiscent of the
soft X-ray excess often observed in some BH accreting sources \citep[][and references therein]{gronkiewicz2020,petrucci2020}.

All models give unabsorbed X-ray flux in the 0.3-10 keV band, within the range 1.0-2.3$ \times 10^{-12}$ erg s$^{-1}$ cm$^{-2}$, as reported in 
Tab.~\ref{tab:spec}. Considering a distance of 9.2 Mpc to NGC 5055, these correspond to an isotropic luminosity in the range 
1.1-2.3$ \times 10^{40}$ erg s$^{-1}$ as listed in Tab.~\ref{tab:spec}. Sources in the luminosity range $10^{40}-10^{41}\ \rm erg/s$ are classified as 
extremely luminous X-ray source (ELX) \citep{devi2007,singha2019}. An ELX resembles the so called  ultraluminous (UL)
spectral state \citep{sutton2013}. UL hard states (dominated by emission from a hot corona) and UL soft states (dominated by 
emission from a cool corona) have been observed \citep{gladstone2009}. {From the data of NGC 5055 X-1 it is not possible to clearly discern whether the source might be in a} UL hard state, 
which is characterized by a hard powerlaw, or in a UL soft state, in which the power law shows a cut-off at lower energies. Given the inferred best-fit electron 
temperatures, the slightly favoured MCD+NTHCOMP model suggests the source to be in a UL soft state.

To illustrate this issue, in Fig.~\ref{fig:model} we plotted the unfolded models for one XMM-Newton observation 
(2007--06--19). From a statistical point of view those models do not differ much, 
but our previous considerations regarding the $N_{\rm H}$ favour the MCD+NTHCOMP model.
Longer observations, extending to hard X-ray energies (e.g. by NuSTAR) are needed to better resolve the 
high energy tail and confirm this interpretation.

\section{Discussions}
\label{sec:results}

\begin{figure}
    \centering
    \includegraphics[trim={0cm 0cm 0cm 0cm},width=0.48\textwidth]{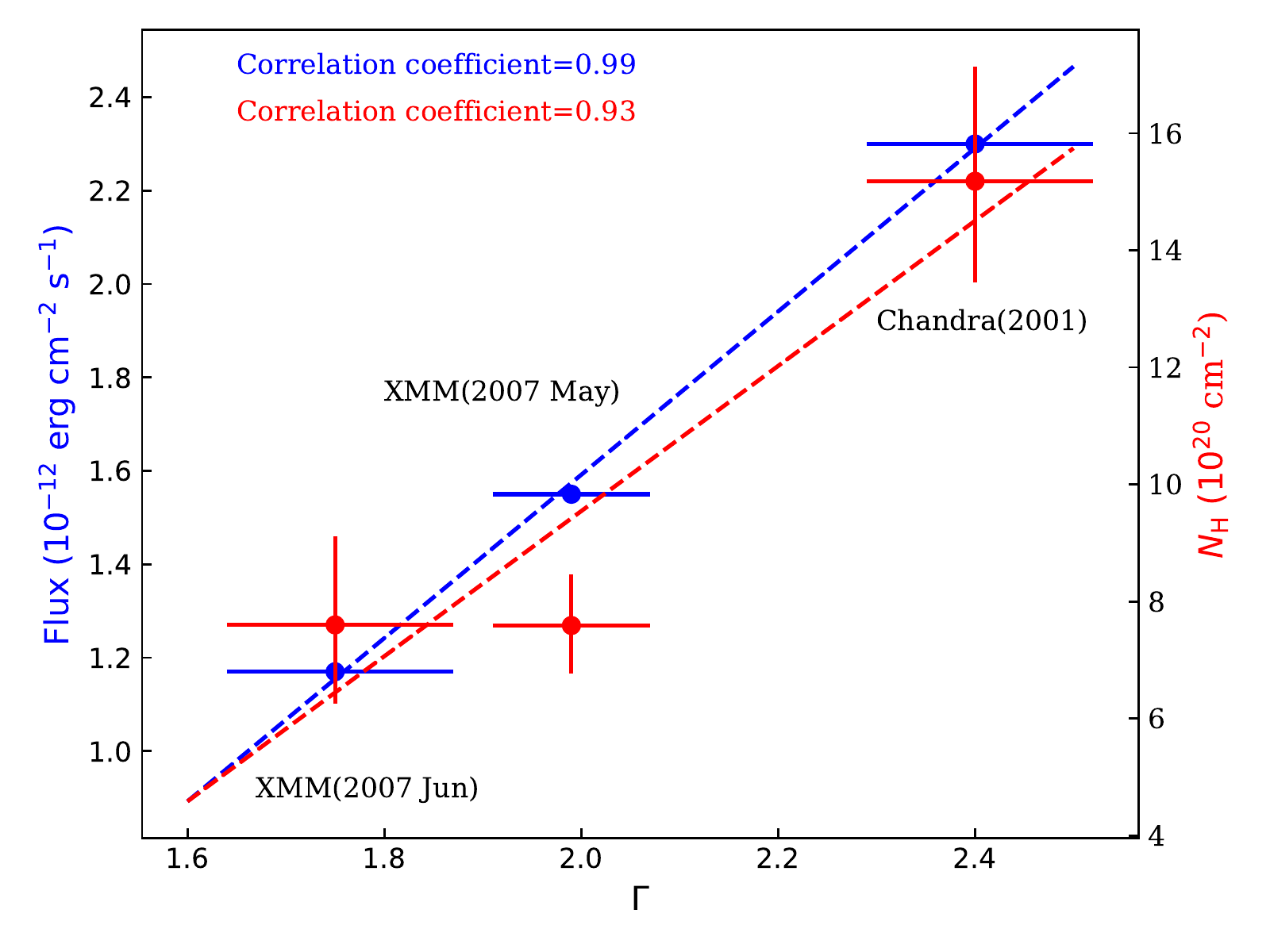}
    \caption{The unabsorbed 0.3-10 keV flux and hydrogen column density versus photon index obtained from MCD+PL model fit. The correlation between 
    $F_{(0.3-10)\rm keV}$ and $\Gamma$  is shown in blue, while correlation between $N_{\rm H}$ and $\Gamma$ is shown red. The data points were fitted 
    with a linear model and the best-fit model is represented by the dashed lines in the plot.}
    \label{fig:gamma}
\end{figure}

Our spectral analysis shows three possible phenomenological models for the spectra of NGC 5055 X-1. The spectral models assume the emission is 
from an accreting BH system. All the three models can describe the data well from a statistical point of view. The results obtained 
from spectral fitting can be used to search for correlations between parameters. In case of the MCD+PL model we found a strong correlation between the 
unabsorbed flux in the range 0.3-10 keV, and the photon index $\Gamma$, with correlation coefficient 0.99, as shown in Fig.~\ref{fig:gamma} (blue points and line). 
The correlation coefficient was measured using the Pearson product-moment.
This correlation shows that the higher the flux the steeper the hard X-ray spectrum, as typically seen in BHXBs above ~0.5-1 percent of the 
Eddington accretion rate \citep{skipper2016}. A similar behaviour was found with correlation coefficient 0.86 in  NGC 1313 X-1, 0.91 in 
NGC 1313 X-2 \citep{feng2006a}, 0.99 in NGC 5204 X-1 \citep{feng2009} and in many other ULXs \citep{feng2006b,kajava2009}.

Following the studies of \cite{gladstone2009} on a large sample of ULXs, we fit the data also with a MCD+PL model. As found by \cite{feng2009} for NGC 5204 X-1 
using the same MCD+PL model, we also found a correlation between $N_{\rm H}$ and  $\Gamma$ with correlation coefficient 0.93, as shown in Fig.~\ref{fig:gamma} (red points 
and line). The correlation shows that as $\Gamma$ increases (simultaneously with the flux) the $N_{\rm H}$ also increases. If this correlation is intrinsic, this might imply 
that an increase of flux leading to strong outflows from the disk, which in turn results in an increase of the hydrogen 
column density of the disk wind. An additional argument for the possible existence of a wind is that the total 
hydrogen column density obtained from all our MCD+PL model fits is at least two times higher than the Galactic value for this source 
(see Sec.~\ref{sec:spec}). This might be the evidence of additional intrinsic absorption 
connected to the source. 
Nevertheless, a confirmation of the presence of a wind in NGC 5055 X-1 can only come from the analysis of high spectral resolution data from the 
Grating spectrometers on \emph{Chandra} and XMM-Newton. We considered available archival data from XMM-Newton reflection gratings (RGS), 
but the exposure time is too short to obtain any significant detection of emission and absorption lines in NGC 5055 X-1. 

In general, the DISKPBB model  gives higher inner disk temperature than the MCD+PL model. 
The important point to note is that we found $T(r)\propto r^{-0.5}$ in all our data sets 
which is flatter than expected for a sub-Eddington accretion disk model. Although the source luminosity changes during different epochs of 
observations, the temperature profile does not respond to such variations 
(the best-fit values of $p$ are constant). \cite{feng2007} performed a similar 
modeling of NGC 1313 X-2 with $p$ as a free parameter, they also obtained high inner disk temperature and $p\sim0.5$. This fact can indicate that the 
disk in ULX sources may be slim, but a definite answer can come only from a systematic fit of a sample of sources using this model.

\begin{figure}
    \centering
    \includegraphics[width=0.48\textwidth]{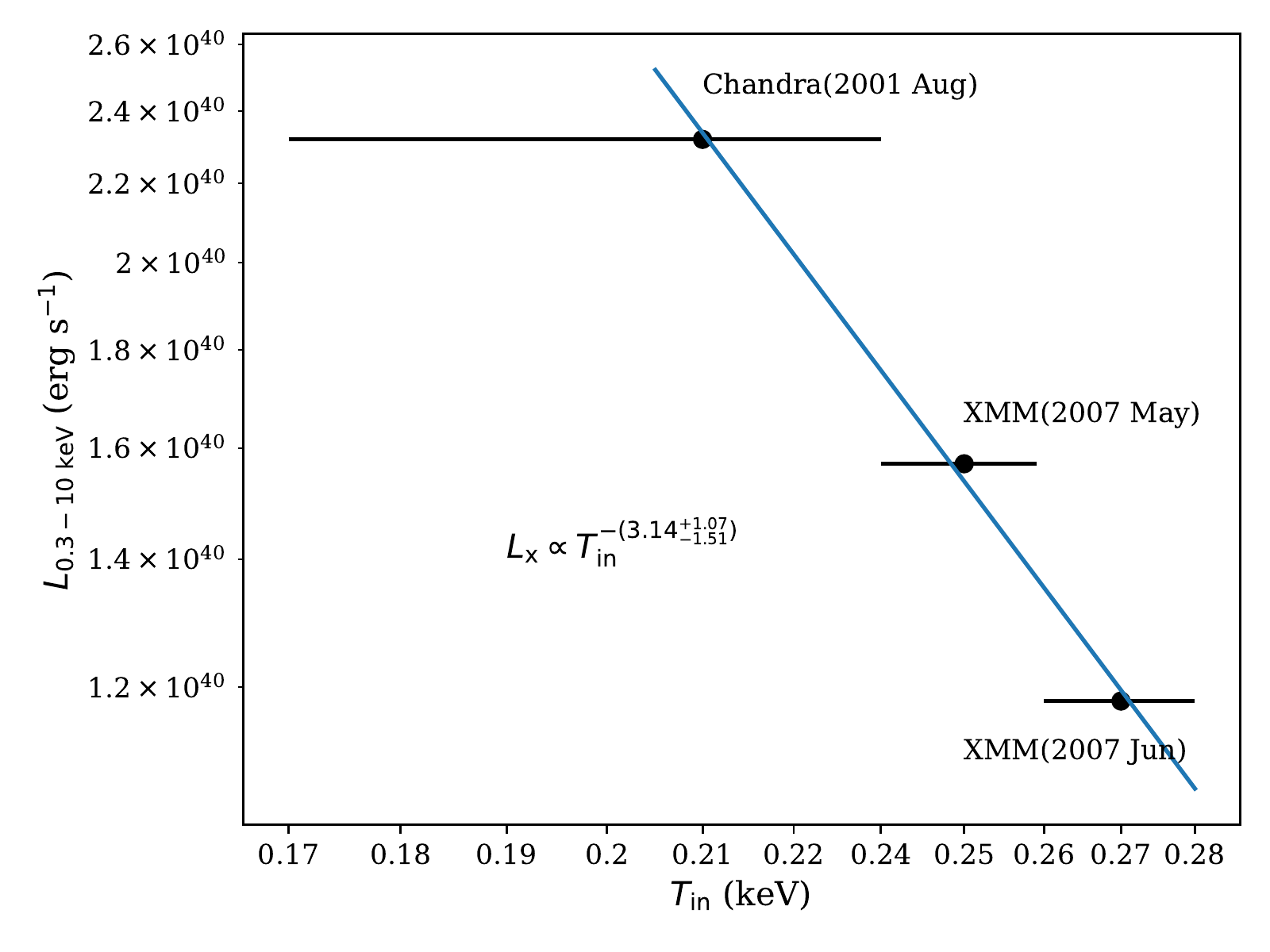}
    \caption{Unabsorbed X-ray luminosity -- inner disk temperature relation inferred from the fit of the MCD+PL model. The blue continuous line shows 
    the best fit to the data which follows the relation 
    $L_{(0.3-10) \rm keV} \propto T_{\rm in}^{-3.14}$.}
    \label{fig:temp1}
\end{figure}

\begin{figure}
    \centering
    \includegraphics[width=0.48\textwidth]{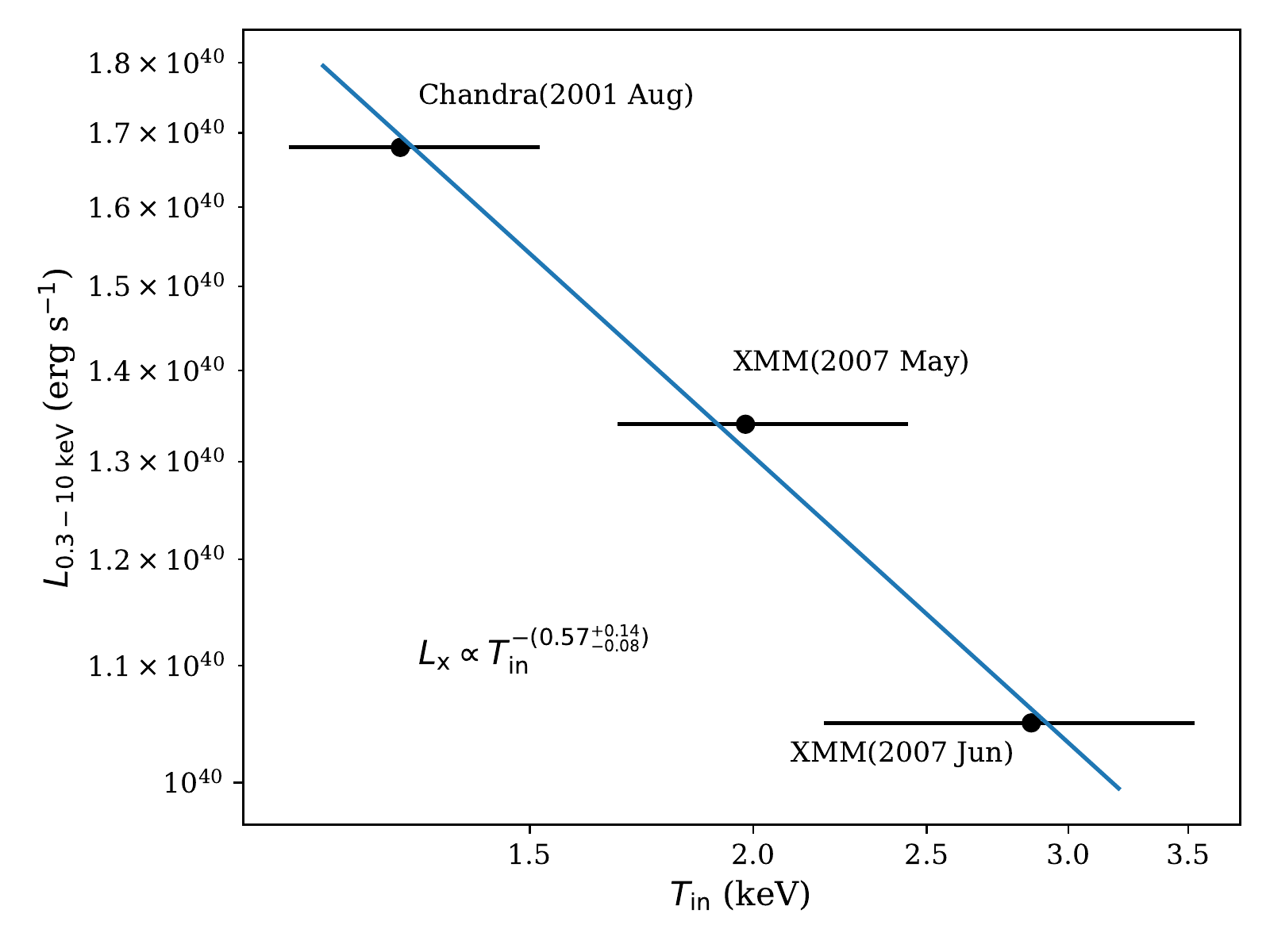}
    \caption{Unabsorbed X-ray luminosity -- inner disk temperature relation inferred from the continuum fitting using slim accretion disk DISKPBB 
    model. The blue continuous line shows the best fit to the data which follows the relation $L\propto T_{\rm in}^{-0.57}$.}
    \label{fig:temp2}
\end{figure}

\begin{figure}
    \centering
    \includegraphics[width=0.48\textwidth]{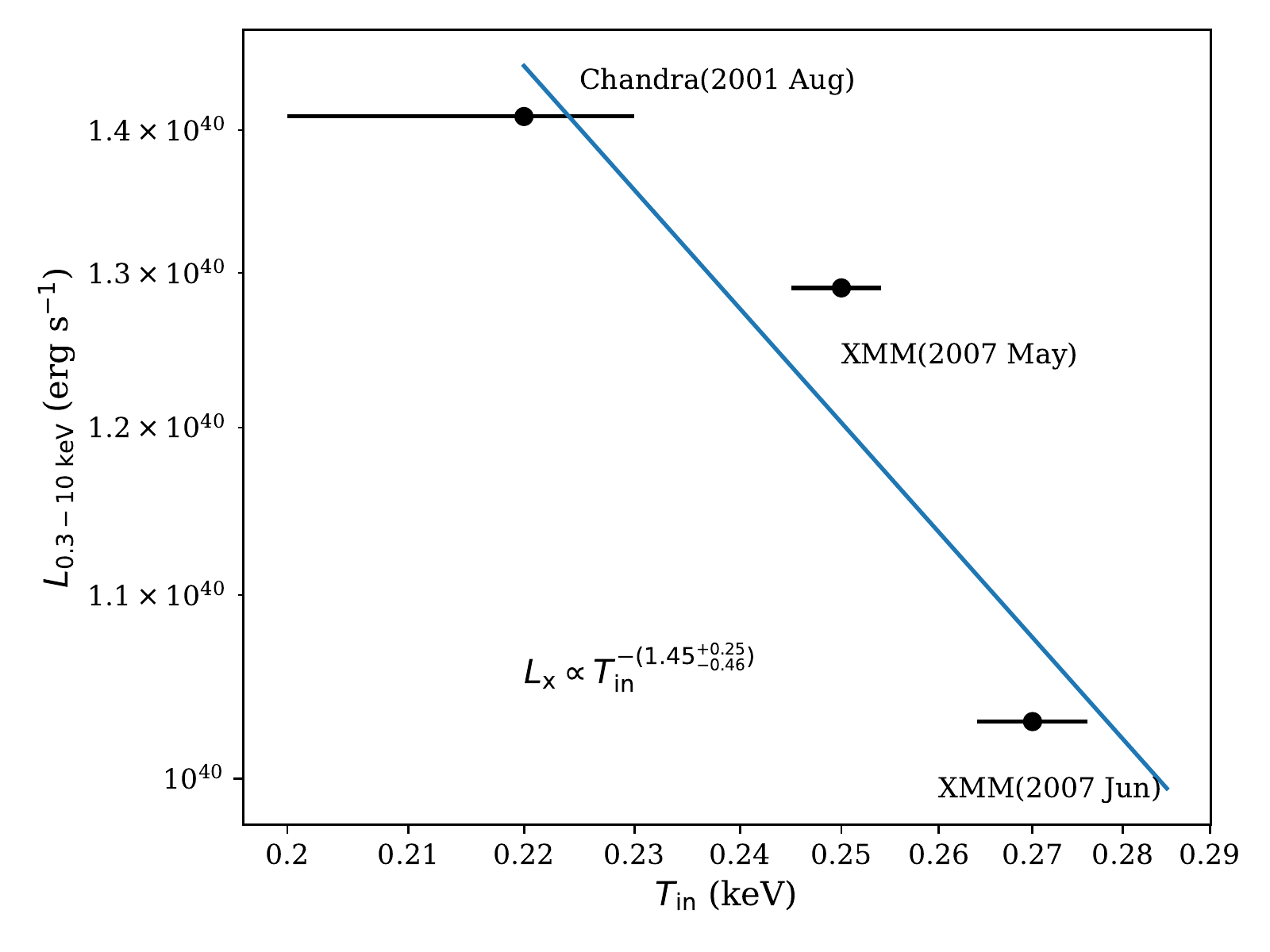}
    \caption{Unabsorbed X-ray luminosity -- inner disk temperature relation inferred from the  fit
    of the MCD+NTHCOMP 
    model. The blue continuous line shows the best fit to the data which follows the relation $L\propto T_{\rm in}^{-1.45}$.}
    \label{fig:temp3}
\end{figure}

The inner temperature $T_{\rm in}$, inferred from MCD$+$PL and MCD$+$NTHCOMP models is relatively low in comparison with the temperature from the 
DISKPBB model. We plot the luminosity versus $T_{\rm in}$ for the three  models in Figs.~\ref{fig:temp1}, \ref{fig:temp2} and \ref{fig:temp3}.
It is clear that NGC 5055 X-1 follows an inverse  relation of luminosity  with $T_{\rm in}$ for all models. The inferred relation is 
$L\propto T_{\rm in}^{-(3.14^{+1.07}_{-1.51})}$ for the MCD+PL model, $L\propto T_{\rm in}^{-(0.57^{+0.14}_{-0.08})}$ for the DISKPBB model, and 
$L\propto T_{\rm in}^{-(1.45^{+0.25}_{-0.46})}$ for the MCD$+$NTHCOMP model. This negative correlation is one of the signatures that the emission 
from the source is geometrically beamed as argued by \cite{king2009}. \cite{feng2007} and \cite{kajava2009} showed that for powerlaw type ULXs the 
soft-bump follows the relation $L\propto T_{\rm in}^{-n}$ where $n\approx3.5$. Among our fits, the MCD+PL follows more closely the predicted 
$n\approx3.5$ trend, although a negative trend is seen in all the other fits. The errors on the inferred relations account for the quite large errors on the temperature. Even considering these errors, the slope of the correlation remains negative, regardless of the model. Of course, the quality of the data and the small number of data points precludes us from obtaining stronger constraints. A longer monitoring is needed to distinguish among the different 
models and confirm these conclusions.


\section{Conclusions}
\label{sec:conc}

We have analyzed the \emph{Chandra} and XMM-Newton observations of the ULX NGC 5055 X-1. The high luminosity of the 
source most likely results from the combination of super-Eddington accretion and geometrical beaming. The source does not show much 
variability in all the three observations. Although the quality of the data does not allow us to draw strong conclusions, our spectral fits hint at interesting results and trends. 

(i) NGC 5055 X-1 mostly emits in soft X-rays in the range of 0.3-3 keV, and the hard X-ray band flux is only a fraction 0.1-0.3 of the 
soft X-ray emission, suggesting a dominant thermal component. Therefore, we tested models of emission from an accretion disk around a BH. 

(ii) The low inner disk temperature ($kT_{\rm in} \sim 0.25$ keV) and the steep power law slope ($\Gamma > 1.5$) obtained in some fits may suggest 
an IMBH. On the other hand the slim disk model DISKPBB provides a good fit, and suggests a temperature emission 
profile $T(r)\propto r^{-0.5}$, which is at odds with that of standard thin disk models \citep{feng2007}.

(iii) All unabsorbed models confirm that NGC 5055 X-1  is intrinsically extremely luminous, reaching 0.3-10 keV luminosity of $2.32 \times 10^{40}$~erg s$^{-1}$.
Our analysis slightly favors the source to be in a soft UL spectral state, but we need better data with longer exposure time to fully confirm 
this conclusion. 

(iv) The flux and $N_{\rm H}$ are positively correlated with the photon index $\Gamma$ for the MCD+PL model. The source jumps from a steep PL 
($\Gamma\approx2.4$) state to a hard PL state ($\Gamma\approx1.8$), while $N_{\rm H}$ varies between  $\sim6-16\times 10^{20}{\rm cm^{-2}}$. This correlation may suggest that the accretion geometry is
disk+corona \citep[][and references therein]{cao2009}. The increasing absorption column density of hydrogen can be interpreted as an outflow 
from the disk.

(v) From a physical point of view, the MCD+NTHCOMP model 
 is slightly more favored, since 
it returns a value of $N_{\rm H}$ consistent with that determined in previous measurements, 
and it allows for a more physical dependence between the two spectral components.

(vi) For all models we find an inverse relation of X-ray luminosity with inner disk temperature. The errors on the temperature are quite large, but regardless of the fitted model, the slope of this correlation is consistent with being negative within the estimated errors. This result strongly suggests that the source is 
geometrically beamed. Therefore, we conclude that a plausible explanation for NGC 5055 X-1 is that the source is accreting at super-Eddington 
luminosity, and is beamed by an optically thick wind, as seen in other high luminosity ULXs \citep{king2009}.

Further simultaneous broad band observations are needed to constrain the full set of physical parameters of NGC 5055 X-1.

\begin{acknowledgements}
We thank the anonymous referee for useful comments that have helped to improve the manuscript.
The authors also thank Aneta Siemiginowska for insights into {\it Chandra} data reduction process. AR was supported  by  Polish National Science 
Center grants No. 2015/17/B/ST9/03422, 2015/18/M/ST9/00541.
BDM acknowledges support from the European Union’s Horizon 2020 research and innovation programme under the Marie Skłodowska-Curie grant agreement 
No. 798726. The work has made use of publicly available data from HEASARC Online Service, Chandra Interactive Analysis of Observations (CIAO) 
developed by Chandra X-ray Center, Harvard \& Smithsonian center for astrophysics (USA), and the XMM-Newton Science Analysis System (SAS) developed 
by European Space Agency (ESA). 
\end{acknowledgements}

%
%

\bibliographystyle{aa}
\bibliography{refs}

\end{document}